\documentclass[useAMS,usenatbib]{mn2e}

\usepackage{amssymb}
\usepackage{aas_macros}
\usepackage{natbib}
\usepackage{times}
\usepackage{graphicx}

\newcommand{\gadget}{{\sc GADGET}}
\newcommand{\flash}{{\sc FLASH}}
\newcommand{\gtsima}{\hbox{$\; \buildrel > \over \sim \;$}}

\title[Numerical overcooling in shocks]{Numerical overcooling in shocks}
\author[Creasey, et al.]  {\parbox[h]{160mm} { 
    Peter Creasey$^{1}$\thanks{E-mail: p.e.creasey@durham.ac.uk},
    Tom Theuns$^{1,2}$, 
Richard G. Bower$^1$ and
    Cedric G. Lacey$^{1}$}
  \vspace{6pt}\\
  $^1$Institute for Computational Cosmology, Department of Physics,
  University of Durham, South Road, Durham, DH1 3LE, UK\\
  $^2$Department of Physics, University of Antwerp, Campus Groenenborger, Groenenborgerlaan 171, B-2020 Antwerp,  Belgium}

\begin{document}

\date{\today}
\pagerange{\pageref{firstpage}--\pageref{lastpage}} \pubyear{2010}

\maketitle

\label{firstpage}

\begin{abstract}
We present a study of cooling in radiative shocks simulated with smoothed particle hydrodynamics (SPH) and adaptive mesh refinement codes. We obtain a similarity solution for a shock-tube problem in the presence of radiative cooling, and test how well the solution is reproduced in  \gadget\ and \flash. Shock broadening governed by the details of the numerical scheme (artificial viscosity or Riemann solvers) leads to potentially significant overcooling in both codes. We interpret our findings in terms of a resolution criterion, and apply it to realistic simulations of cosmological accretion shocks onto galaxy haloes, cold accretion and thermal feedback from supernovae or active galactic nuclei. To avoid numerical overcooling of accretion shocks onto haloes that should develop a hot corona requires a particle or cell mass resolution of $10^6 M_\odot$, which is within reach of current state-of-the-art simulations. 
At this mass resolution, thermal feedback in the interstellar medium of a galaxy requires temperatures of supernova or AGN driven bubbles to be in excess of $10^7$~K at densities of $n_{\rm H}=1.0 \, {\rm cm}^{-3}$, in order to avoid spurious suppression of the feedback by numerical overcooling.
\end{abstract}


\section{Introduction}
Radiative cooling and shocks are two important ingredients in galaxy formation theory \citep{White_and_Rees_78}. Whilst most codes used in astrophysics have facilities for handling both of these, the scales at which these operate and interact are challenging. We will start with a short tour of the processes and of the numerical codes we will use to simulate them. We describe a 1-dimensional model problem of a radiatively cooling shock with an analytic solution which we model with two popular codes in astrophysical simulations, \flash\, \citep{Fryxell00} an adaptive mesh refinement (AMR) code, and \gadget\, \citep{Springel_05}, a smoothed particle hydrodynamics code(SPH; \citealp{Gingold_and_Monaghan_77,Lucy_77}). The results of our simulations give appropriate criteria with which we can analyse the efficacy of our numerical schemes in a wide variety of astrophysical environments. We also investigate the mitigating factors such as the ratio of the cooling to dynamical times, which may enable a simulation to give approximately correct results when otherwise we would consider there to be insufficient resolution.

There has been considerable discussion of the treatment of discontinuities in SPH \citep{Price2008,Read2010}, motivated by problems illustrated by \cite{Agertz2007}, but the issues highlighted in this paper are of a different nature. Those papers focus on spurious forces introduced by the SPH scheme whilst we focus on evaluating the errors introduced as we approach the resolution limit, which are to some extent unavoidable. 

\subsection{Astrophysical shocks}
The science of astrophysics is an ideal domain for the investigation of shock fronts on a variety of scales. Stellar winds form shocks as they push in to the interstellar medium. On larger scales SNe form very high Mach number shocks as they plough into the surrounding gas and form remnants. On larger scales still, galactic winds, driven by starbursts and active galactic nuclei (AGN), shock against the inter galactic medium (IGM). In the context of galaxy formation we can also consider accretion shocks, where gravitationally accelerated infalling gas shocks to form a hot corona in the dark matter potential well.

Of particular interest to us in this paper are radiatively cooling shocks. To an extent, all the aforementioned shocks have radiative cooling, however the cosmological accretion shocks and SNe are particularly topical. In galaxy formation simulations the SNe at early times form remnants well below the resolution of current simulations and need to be modelled with subgrid physics (see \citealp{Kay_et_al_02} for a review of feedback methods). The cooling of the hot gas causes a transition from a thermally-driven to a momentum-driven phase, losing a significant fraction of the SNe energy. A similar transition is thought to occur in the thermal to momentum transition of winds powered by an AGN \citep{Booth_and_Schaye_09}.

Cooling in accretion shocks may affect the fuelling of star formation in the host galaxy. If the gas is shocked to too high a temperature it will not cool over a Hubble time, preventing star formation (though non-spherical geometries may allow the gas to compress first and thus cool faster, see e.g. \citealp{Birnboim_and_Dekel_03}). In a cosmological simulation, however, the resolution around these cooling regions may be so coarse as to resolve these cooling regions with only a few particles. In this paper we intend to probe the effect of limited numerical resolution in these cases, and how these may affect the outcome of the simulation.

\subsection{Physical shock fronts}
Before we concentrate on the numerical aspect of cooling in shocks, we begin by briefly considering the processes that occur in a real physical shock front. A shock front is a region where one of the usually conserved fluid properties, entropy, is allowed to change. It is worth considering why such a property is otherwise treated as a constant, and why shocks are a special case.

In the kinetic theory of gases, a gas is described as a large number of particles (e.g. atoms, molecules, ions) in constant random motion. The frequency of collisions defines a timescale, and also a typical length between collisions, the mean free path. If all processes acting on the gas happen on timescales much greater than the time between collisions, then the classical theory of adiabatics tells us that there will be another conserved property which, for an ideal gas, is $p/\rho^{\gamma}$. Here, $p$, $\rho$ and $\gamma$ are the pressure, density and adiabatic index, respectively. This property is a function of the entropy, and in astrophysics is often used as a proxy.

It is worth recalling that processes which change the fluid entropy (e.g. shocks, radiative absorption, thermal conduction) will occur on timescales on the order of, or shorter than, the period between collisions (or over lengths on the order of the collision length). Mechanisms which heat the gas on slower time scales will be \emph{adiabatic} processes, and will alter the thermal energy with only very small increases in entropy\footnote{One can of course construct systems in which the time scale of interest is long enough such that viscosity, thermal diffusion, etc. dominate the large scales too. These problems, however, have low Reynolds and P\'eclet numbers respectively, and are the exception rather than the norm in computational astrophysics}.

Now we come to shock fronts. A canonical example of a shock front would be a 1 dimensional system where the upstream fluid travels supersonically with respect to the down wind fluid (i.e faster than the thermal velocities of the particles), until it reaches the shock, where the majority of its mechanical energy (the bulk motion of the particles) is converted into thermal energy. This happens because the pairs of particles that collide can have very different velocities: particles in the shock front change their energy on a timescale on the order of the collision time between a pair of up and down wind particles, which is much shorter than that between two down wind, or two upwind, particles. From this description we immediately see that physical shocks must occur over length scales on the order of the mean free path, which is usually much smaller than other physical length scales in the problem.

The mean free path ($\Delta x$) depends upon the number density of particles ($n$) and their collisional cross section ($\sigma$), as
\begin{equation}
\Delta x = \frac{1}{n\sigma}\,.
\end{equation}
In the case of a partially or fully ionized gas, particles may interact on a shorter length scale \citep{Zeldovich_67}, that of the plasma skin depth/plasma oscillation length
\begin{equation}
\Delta x = c  \left( \frac{4\pi n_e e^2}{m_e}\right)^{-1/2} \approx
10^6 \left( \frac{n_e}{1\, {\rm cm}^{-3}} \right)^{-1/2} \, \mathrm{cm}\,.
\end{equation}
Since the particles are not interacting via Coulomb collisions this is known as a \lq collisionless shock\rq; the mechanism of interaction is the plasma oscillation (coupling together charged particles). Care must be taken, however, as the post-shock gas may be out of thermal and ionizational equilibrium for the problem in question (something that would not be a concern if the collision length is small), making these cases challenging to simulate.

The trapping of relativistic ions between magnetic fields in the up and down-stream phases and subsequent acceleration is also believed to be the origin of the power law spectrum of high-energy cosmic rays, a Fermi acceleration process. 

Finally, we should complete this discussion by mentioning turbulence as a source of entropy. In general the bulk oscillations of fluids will occur on scales much larger than the mean free path and is thus unable to change the entropy. Transfer of spectral energy to shorter wavelengths, however, implies that eventually bulk oscillations reach the scale of the mean free path and will be dissipated into thermal energy \citep{Kolmogorov_41}.

\subsection{Shocks in simulations, artificial viscosity}
Now let us consider shocks in simulations. Almost exclusively, the resolution of simulations will be much coarser than a physical shock width. This is not necessarily a problem, however, as the bulk properties of the post-shock gas may be deduced from the conservation of energy and momentum, and the assumption that the shock process does not produce oscillations on scales much larger than the mean free path.

In this paper we will contrast two schemes for numerical hydrodynamics that are popular in cosmology: SPH and AMR. Smoothed Particle Hydrodynamics (SPH; \cite{Gingold_and_Monaghan_77,Lucy_77}, see \cite{Monaghan_05} and \cite{Springel_05} for recent reviews) is a (pseudo) Lagrangian scheme in which the fluid is represented by a set of particles that move along with the flow. In this paper we will illustrate the behaviour of SPH using the \gadget\, SPH implementation of \cite{Springel_05}. Adaptive Mesh Refinement (AMR) follows how fluid flows across a (stationary) computational mesh, whose cell size may be locally \lq refined\rq\ or \lq de-refined\rq\, based on some criterion. In this paper we use the \flash\, code, a block-structured AMR implementation by \cite{Fryxell00}.

The physical process of kinetic energy dissipation by particle collisions is represented in the continuum approximation by a viscous term in the Navier-Stokes equations,
\begin{equation}
\label{eq:nav_stokes}
\frac{\partial}{\partial t} (\rho v_i) + \frac{\partial}{\partial x_j} (\rho v_i v_j) = \frac{\partial p}{\partial x_i} +  \frac{\partial}{\partial x_j} \sigma_{ij}\,,
\end{equation}
where
\begin{equation}
\sigma_{ij} = \eta \left(\frac{\partial v_i}{\partial x_j} +\frac{\partial v_j}{\partial x_i} -\frac{2}{3} \frac{\partial v_k}{\partial x_k}  \delta_{ij} \right) + \zeta \frac{\partial v_k}{\partial x_k} \delta_{ij}\,,
\end{equation}
is the viscous stress tensor and $\eta$ and $\zeta$ are known as the shear and bulk viscosity coefficients, respectively.  These coefficients can be measured for real fluids, however in most astrophysical flows they are so small that the viscous term is insignificant outside of shocks (i.e. the flows have high Reynolds number).

The variant of SPH used in this paper handles shocks with a prescription known as artificial viscosity (although Godunov type methods for SPH also exist, see \citealp{Inutsuka_94}). Artificial viscosity was originally developed for grid codes \citep{Vonneumann_Richtmyer50}, and use the bulk viscosity term in Eq.~(\ref{eq:nav_stokes}), however, with the coefficient raised by several orders of magnitude. These larger values prevent the shocks generating large unphysical oscillations due to the coarseness of the sampling (see the numerical stability criterion of \citealp{Lax_Friedrichs_71}). In SPH they also fulfil a second role of preventing particle interpenetration (see \citealp{Bate_95} for a thorough discussion). A number of artificial viscosity prescriptions are in use, the most common being that of Monaghan \& Balsara \citep{Balsara_95}, a Lax-Friedrichs style viscosity that is turned on for compressing flows. The implementation in our version of Gadget is based on signal velocities \citep{Monaghan_97}.

In mesh codes shocks can be treated with artificial viscosity but more commonly a conservative Riemann solver (based upon Godunov's scheme, \citealt{Godunov_64}) is used. Riemann solvers (see e.g the HLL solver, \citealp{Harten_83}) give exact solutions in 1d or planar shock problems with homogeneous pre- and post-shock fluids, but are somewhat diffusive in other cases. They are still the preferred method for grid codes, however, and the default used in \flash\ is a directionally split Riemann solver \citep{Colella_Woodward_84}. Oscillations near the discontinuities are controlled with a monotonicity constraint.

\subsection{Radiative cooling}
Radiative cooling is an essential ingredient in galaxy formation as it is the process which allows the baryons in dark matter halos to dissipate thermal energy and thus collapse to form galaxies. Multiple cooling mechanisms are important in the astrophysical domain, however in this paper we will primarily be interested in collisional line cooling and at higher temperatures, thermal bremsstrahlung. The evolution of the specific thermal energy, $u$, due to cooling can be written as
\begin{equation}
\left. \rho \dot{u} \right|_\Lambda = -\Lambda(T; Z) n^2\,, \label{eq:cool}
\end{equation}
where $\rho$ is the density, $T$ the temperature, $Z$ the metallicity and $n$ the particle number density (for brevity we will subsequently refer to the radiative component of the specific cooling rate $\left.  \dot{u} \right|_\Lambda$ as $\dot{u}_\Lambda$). When baryon-photon interactions with the CMB and an ionizing background are important we have followed the prescriptions of \cite{Wiersma_Schaye_and_Smith_09} (see also Fig. \ref{fig:gimiccool}, below).

The implementation of cooling in our versions of \gadget\ and \flash\ is performed by an adaptive time step integration over each cell/particle. The effects of cooling are included in the hydrodynamic solver by \emph{operator-splitting}, i.e. the separation of the two processes $A$ (radiative cooling), and $B$ (shock heating)  into individual steps,
\begin{eqnarray}
\dot{X} &=& \left(A + B \right) X  \\
X_{t+\Delta t} - X_t &\approx& A(\Delta t) B(\Delta t ) X \, , 
\end{eqnarray}
where the errors on the latter term on the order of the time step $\Delta t$ depend upon the commutator $[A,B]$. Since the physical process of shock heating should occur over a much shorter time scale than that of radiative cooling we can justify this being zero. The numerical implementation of shock heating will of course take a longer time scale and thus would interact with the cooling if operator splitting was not introduced, however there is no physical motivation to prefer such a scheme in this case. 

\section{Radiatively Cooling shocks, a model problem}\label{sec:RadCoolingSetup}
A typical test problem in numerical hydrodynamics is that of the formation of a 1-dimensional shock in a 
\lq test tube\rq.
In one form of this problem a tube is initialised with gas of constant polytropic index $\gamma$, the left and right halves converging with opposing velocities $v_0$ and $-v_0$. For a sufficiently high velocity $v_0$ a shock will form, creating a downstream region with higher temperature, pressure and density. This problem has a similarity solution for constant $\gamma$. In our set-up the gas is allowed to cool radiatively, and the downstream region can then cool to form a dense post-shock region. The initial conditions are thus

\begin{eqnarray}
\rho(x,t_0) &=& \rho_0 \\
p(x,t_0) &=& p_0  \\
T(x,t_0) &=& T_0 \\
v(x,t_0) &=& \left\{ \begin{array}{cc} 
v_0 , & x<0  \\
-v_0 ,&  x>0 \end{array} \right.\,
\end{eqnarray}
We note that the symmetry of this problem makes it equivalent to the wall shock (where there is an immovable boundary at $x=0$, \citealt{Monaghan_97}). In order to minimise the amount of modification in our SPH code we chose to set up the symmetric problem rather than implement an immovable boundary.
\subsection{Similarity solution for a radiative 1D shock}

\begin{figure*}
\centering
\includegraphics[width=2\columnwidth]{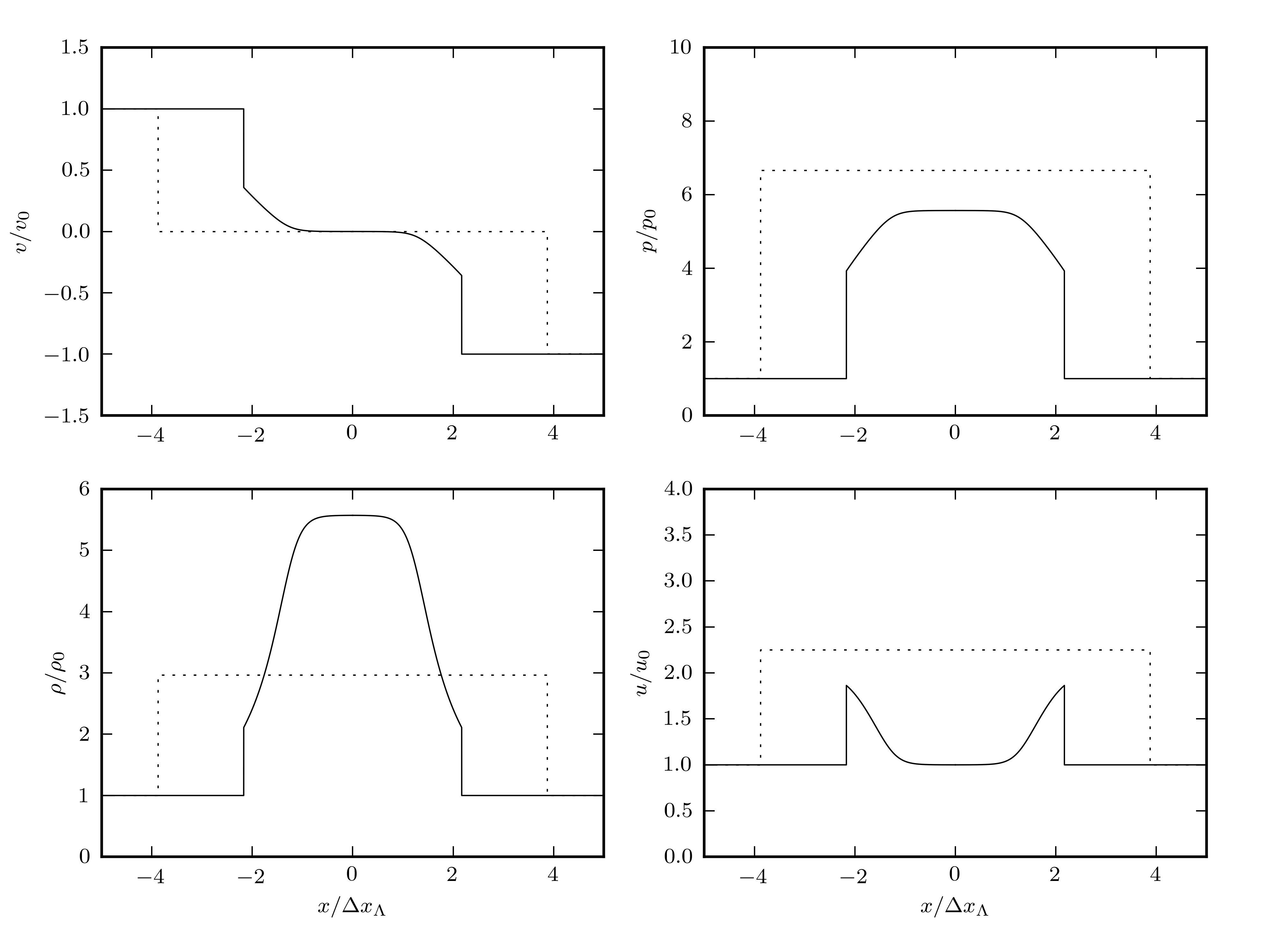}
\caption{\emph{Solid lines} represent the analytic solution for the colliding gas problem discussed in Section \ref{sec:RadCoolingSetup} when cooling is included. Incoming gas from the left and right shocks and then compresses and cools to form a cold dense region in the center. For the example shown, the Mach number of the upstream gas is $1.5$ w.r.t. the cold, central gas and the time is $5.1 \, \Delta x_\Lambda/c_0$ (see Eq.~(\ref{eq:coolinglength})). For comparison, \emph{dashed lines} show the solution without cooling at the corresponding time. At early times (i.e. $t \lesssim \Delta x_\Lambda / c_0$) the cooling profile is not of given form, as it has not had sufficient time to reach a stationary state (details of the similarity solution for cooling through a shock can be found in appendix \ref{sec:Piecewise}).}
\label{fig:analytic}
\end{figure*}

If the temperature dependence of the cooling rate is sufficiently simple, then this 1 dimensional shock problem has a similarity solution even in the presence of cooling. Such is the case for a cooling function which is a piecewise linear function of the temperature, such that the rate of radiative cooling, $\dot{u}|_\Lambda$, of the specific energy $u$, is given by a \lq cooling spike\rq\,:

\begin{equation}
\left. \rho \dot{u} \right|_\Lambda = \left\{ \begin{array}{cc} 
0 , & T< T_0  \\
-\Lambda n^2 (T-T_0)/T_0 ,&  T_0 \leq T \leq \frac{1}{2} (T_1 + T_0) \\
-\Lambda n^2 (T_1 - T)/T_0 , & \frac{1}{2} (T_1 + T_0)  \leq T \leq T_1 \\
0 , &  T_1 \leq T  \end{array} \right.
\label{eq:cooling}
\end{equation}
where $\Lambda$ is a positive constant and cooling is maximum at $T = \frac{1}{2} (T_1 + T_0)$. For
simplicity, in all simulations discussed below we initialise the temperature to $T_0$ (where the cooling vanishes), so the initial gas is not cooling. 

The gas is chosen to be pure atomic hydrogen, i.e.
\begin{eqnarray}
\gamma &=& \frac{5}{3} \\
\rho &=& m_p n \\
p &=& n k_B T\,.
\end{eqnarray}

For comparison the reader should see the simulations of \citet{Hutchings_Thomas_00} who used a more realistic astrophysical cooling function, at the expense of not having an analytic solution. For the cooling post-shock region we find analytic solutions in a similarity variable $\lambda \equiv \rho / \rho_0$ of the form (see Appendix \ref{sec:Piecewise} for details)

\begin{eqnarray*}
T / T_0 &=& ( (a+1) \lambda^{-1} - a \lambda^{-2} ) \\
x - x_0 &=& \frac{-v_s k_B T_0 }{(\gamma - 1) \Lambda
  n_0}\left[\frac{\gamma-a}{a-1} \log (1 - \lambda^{-1}) + \right. \\
& &  \frac{1-a\gamma}{(a-1)a^2} \log (1 - a\lambda^{-1}) \\
& & \left. - \frac{a+1}{a} \lambda^{-1} - \frac{\gamma+1}{2}
  \lambda^{-2}  \right] \\
a & \equiv & \frac{\rho_0 v_s^2}{p_0}.
\end{eqnarray*}

The value of the shock velocity $v_s$ and the final density in the cold, post-shock region $\rho_0$ can be found by imposing conservation of mass and momentum (see Appendix \ref{sec:RHShock}). The solution is shown in Figure \ref{fig:analytic}.

\subsection{Shock stability}
\citet{Chevalier_Imamura_82} find that positive increasing linear cooling functions produce stable shocks. Applying this to the cooling function in Eq.~(\ref{eq:cooling}) we see that we have stable shocks provided the post shock temperature $T_s < \frac{1}{2} (T_1 + T_0)$ or $T_s > T_1$, which is the case for all the shocks we study later (we define the shock temperature $T_s$ as the temperature immediately after the shock, which is computed in Appendix \ref{sec:RHShock}). If the gas is shocked to $\frac{1}{2} (T_1 + T_0) < T_s < T_1$ then the shock may be unstable as the cooling function has a negative slope, $\partial_T \left( - \rho \dot{u}_\Lambda \right) < 0$. Intuitively this can be understood in terms of the length of the cooling region: if the length increases the shock velocity will be higher, causing the post shock gas to be hotter, which increases the cooling time, which feeds back into a longer cooling region.

\subsection{Numerical solution}
\subsubsection{Initial conditions}
The similarity solution is described with two (dimensionless) parameters. The first is the ratio of the upper to the lower temperature in the cooling spike, which we will set to $20$, i.e. $T_1 = 20 T_0$. This is motivated by the temperature dependence of the radiative cooling function of an astrophysical plasma (see also Fig. \ref{fig:gimiccool}), where individual elements contribute significantly to the cooling over approximately a 1 dex range in temperature.

The second parameter is the Mach number of the shock, which we will quote in the rest frame of the problem (rather than the rest-frame of the post-shock gas, for example), 
\begin{equation}
\mathcal{M} \equiv \frac{v_0}{c_0}\,,
\end{equation}
where $c_0\equiv (\gamma p_0/\rho_0)^{1/2}$ is the upstream sound speed. Our tests are performed at $\mathcal{M} =4.70$ and $\mathcal{M} =6.04$. The former has been chosen such that the shock reaches a temperature somewhat below the maximum of the cooling function, $(T_1 + T_0)/2$ (where the shock will be stable) and the latter such that the shock reaches a temperature somewhat above $T_1$ (where there is no cooling). 

We plot positions in units of the cooling length,
\begin{equation}
\Delta x_\Lambda \equiv \frac{k_B T_0 c_0}{ \Lambda n_0}\,.
\label{eq:coolinglength}
\end{equation}
Similarly we express times in units of $\Delta x_\Lambda/c_0$. As observed in \citet{Monaghan_97}, numerical schemes (including both SPH and AMR) usually produce a transient unphysical thermal bump at $t=0$ when there is no post-shock region. To avoid contamination by this transient we run our simulation for a time $14.2 \Delta x_\Lambda / c_0$ and $7.1 \Delta x_\Lambda / c_0$ for the $\mathcal{M} = 4.7, 6.04$ shocks respectively (i.e. we simulate for several sound crossing times of the cooling region, to make sure it is in a stationary state).

For our SPH simulations we set up a long box, periodic along all boundaries. The particle mass is chosen to be
\begin{equation}
m_{\rm SPH} = \rho_0 \cdot (0.3 \Delta x_\Lambda)^3\,,
\end{equation}
(i.e a mean inter-particle spacing of $0.3 \Delta x_\Lambda$). We note that this setup creates a rarefaction wave that propagates inwards from the far edges of the computational volume (due to the discontinuity on this boundary), and thus we need a box long enough such that these rarefactions do not reach our domain of interest in the simulation time. The particles were set up with a `glass' distribution \citep{White_94} to minimise relaxation effects in the pre-shock fluid (the SPH kernel allows a cubic lattice arrangement of particles to slightly reduce its density, and hence release some thermal energy, by relaxing to a glass like state). We also raised the level of the bulk artificial viscosity constant, $\alpha$, to 3 (from 1, see \citealt{Springel_05} for a complete description of the artificial viscosity prescription). We found this to be necessary to prevent ringing and the appearance of large scatter in the entropy of SPH particles in the post-shock region (see also \citealt{Abel_10}).

For the AMR simulations we again set up a long box with cell spacing $ 0.3 \Delta x_\Lambda$, with periodic boundaries in the $y$ and $z$ directions and inflowing gas along the (long) $x$ axis. No refinement was allowed, effectively making this a uniform Eulerian mesh.

We considered allowing an alternative refinement criterion, however the standard \flash\, refinement schemes will refine a shock to the maximum allowed level (since it contains a discontinuity), reducing it to the uniform mesh case. We refer the reader to the dashed lines in Figs. \ref{fig:lowMach} \& \ref{fig:highMach} to compare resolutions.

We note that the use of inflowing boundary conditions in \flash\ allowed us to avoid the rarefaction waves we created in SPH, and thus we could use a much shorter box (by a factor 10). To set the scene we begin by looking at shocks in the absence of cooling.

\subsubsection{Test without cooling}
\begin{figure}
\centering
\includegraphics[width=\columnwidth]{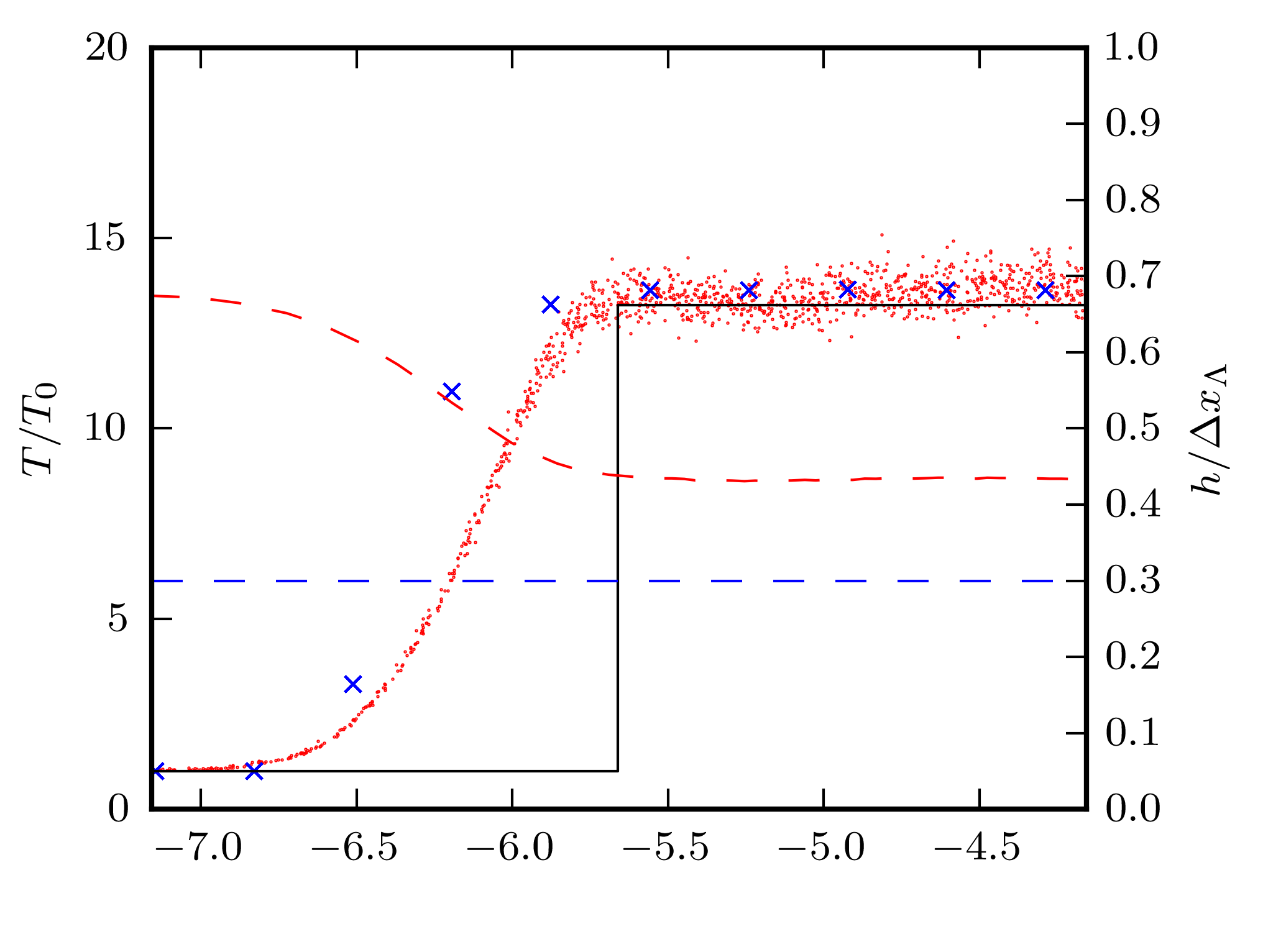}
\includegraphics[width=\columnwidth]{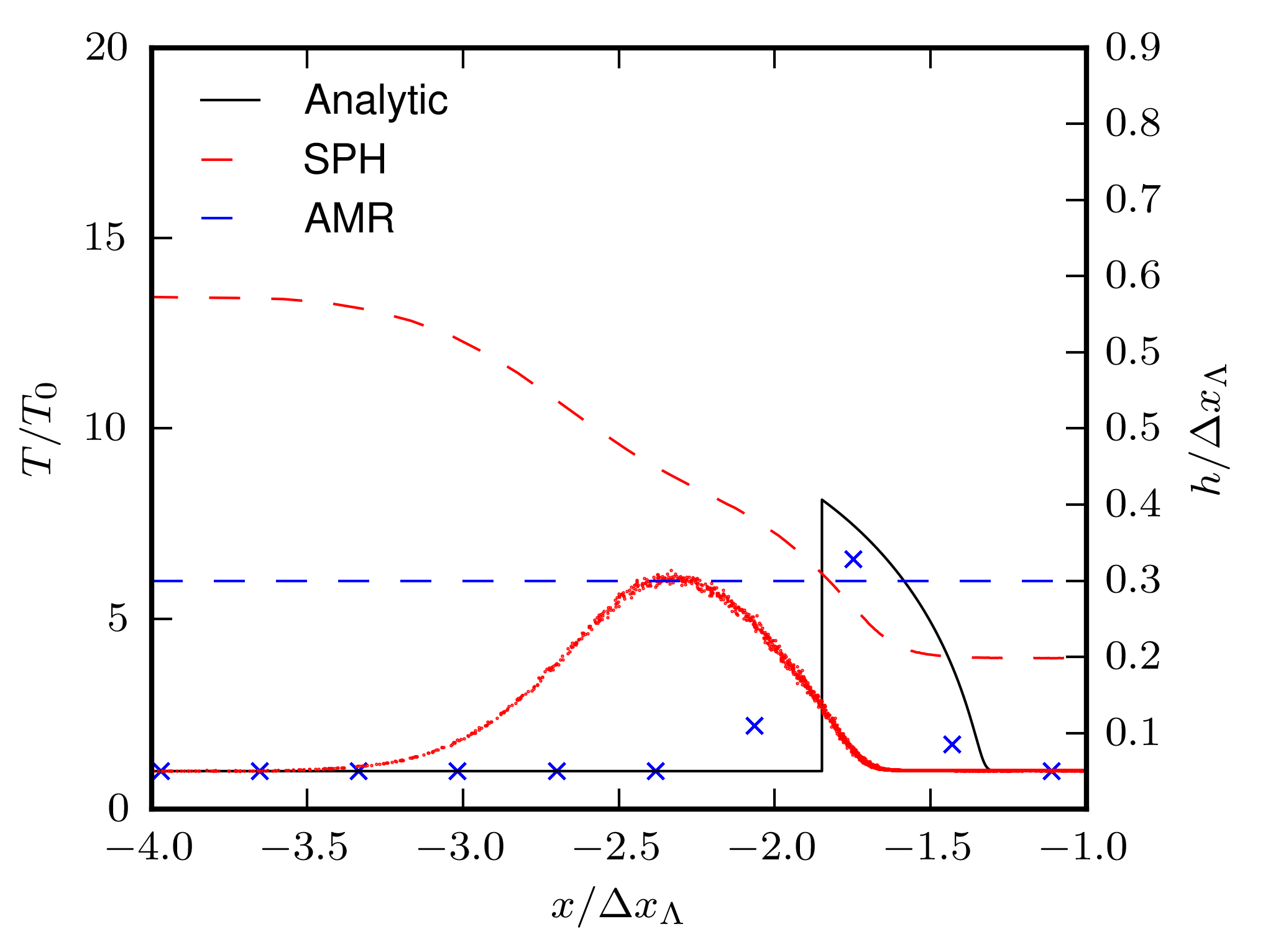}
\caption{{\it Upper panel} plots the temperature in a Mach $\mathcal{M}=4.7$ shock without cooling. Each \emph{blue
  cross} represents a column of \flash\ cells (the tube is orthogonal to the mesh), each \emph{red point} represents a \gadget\ 
  particle, the {\it black line} is the analytic solution. \emph{Red dashes} denote the smoothing lengths of the \gadget\ particles, \emph{blue dashes} the separation of \flash\, cells (right axes). Incoming gas from the left (and right, not shown) collides to form a
  homogeneous hot, rarefied region in the centre. As expected, both codes reproduce the correct profile relatively well. The
  shock is seen to be spread over several cells (\flash) or smoothing lengths (\gadget). {\it Lower panel} as for top panel but including cooling. The analytical solution shows that the gas shocks to a lower temperature (due to the smaller difference between the incoming gas velocity and the shock velocity), followed by a \lq cooling tail\rq in the post-shock region. When simulated using \gadget\, SPH particles shock in several steps before reaching their maximum temperature. As they do so, particles cool to some extent in the smoothed shock and hence
reach a lower maximum temperature than the analytical solution (the SPH shock is also offset to the left of the analytic shock). In the \flash\ run gas gets shocked to higher temperatures, closer to the analytical solution.  Note that as the gas gets compressed the downstream SPH smoothing length is smaller than the \flash\ cell size.}
\label{fig:lowMach}
\end{figure}
\begin{figure}
\centering
\includegraphics[width=\columnwidth]{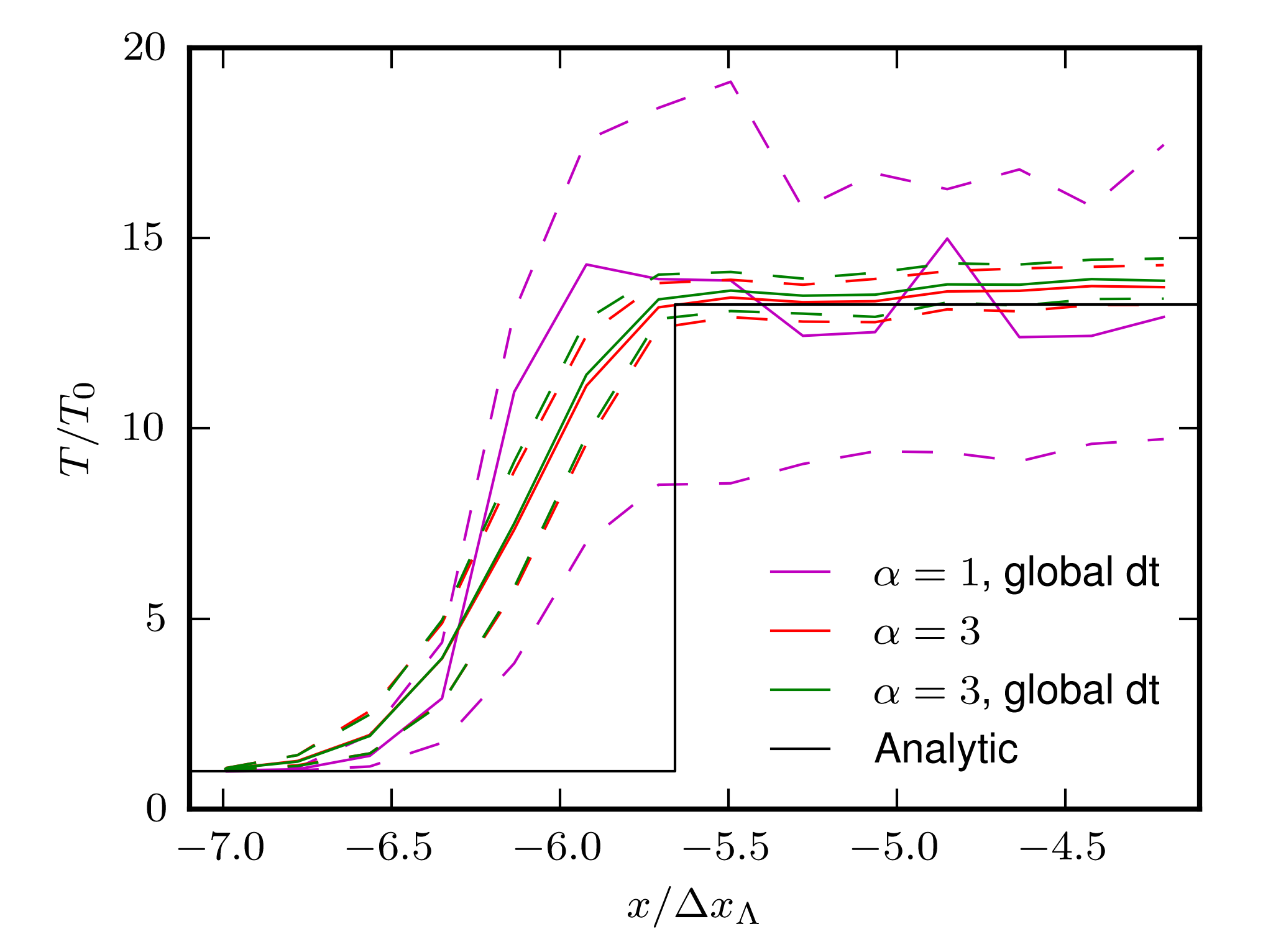}
\caption{Comparison of the effects of altering the viscosity and timestep on the shock from the upper panel of Fig.~\ref{fig:lowMach}. \emph{Red lines} show the higher viscosity ($\alpha=3$) scheme with adaptive time steps, \emph{green lines} the same viscosity but a global minimum timestep (set to the minimum Courant step of all particles) and \emph{purple lines} the global minimum timestep but with the default viscosity ($\alpha=1$), {\emph black line} the analytic solution. \emph{Dashed lines} show the 10th and 90th percentiles.}
\label{fig:avEffects}
\end{figure}

The test problems in the absence of cooling are compared in the upper panels of Fig.~\ref{fig:lowMach} and \ref{fig:highMach} ($\mathcal{M}=4.7,6.04$ respectively). Provided we use the higher than usual value of the artificial viscosity ($\alpha=3$) in \gadget , both the SPH and AMR codes handle this shock well (as expected), with the shock smeared out over a few times the resolution length $h$ in SPH, and a few cells in \flash. At \gadget 's default value for the artificial viscosity ($\alpha=1$) we find that this is too high a Mach number shock to be handled (we do however return to the original value when we study the lower Mach number shocks in section \ref{sec:resCrit}). In Fig.~\ref{fig:avEffects} we tested both altering the value of the artificial viscosity and adjusting the maximum time step (between \gadget 's default adaptive scheme and a global minimum Courant step applied to all particles). The higher value of artificial viscosity was found to significantly reduce the scatter in the post shock thermal energies, superior to a reduction in the global time step. We would, however, expect that at very high Mach number shocks a more conservative time step would be required.

\subsubsection{Test with cooling}

\begin{figure}
\centering
\includegraphics[width=\columnwidth]{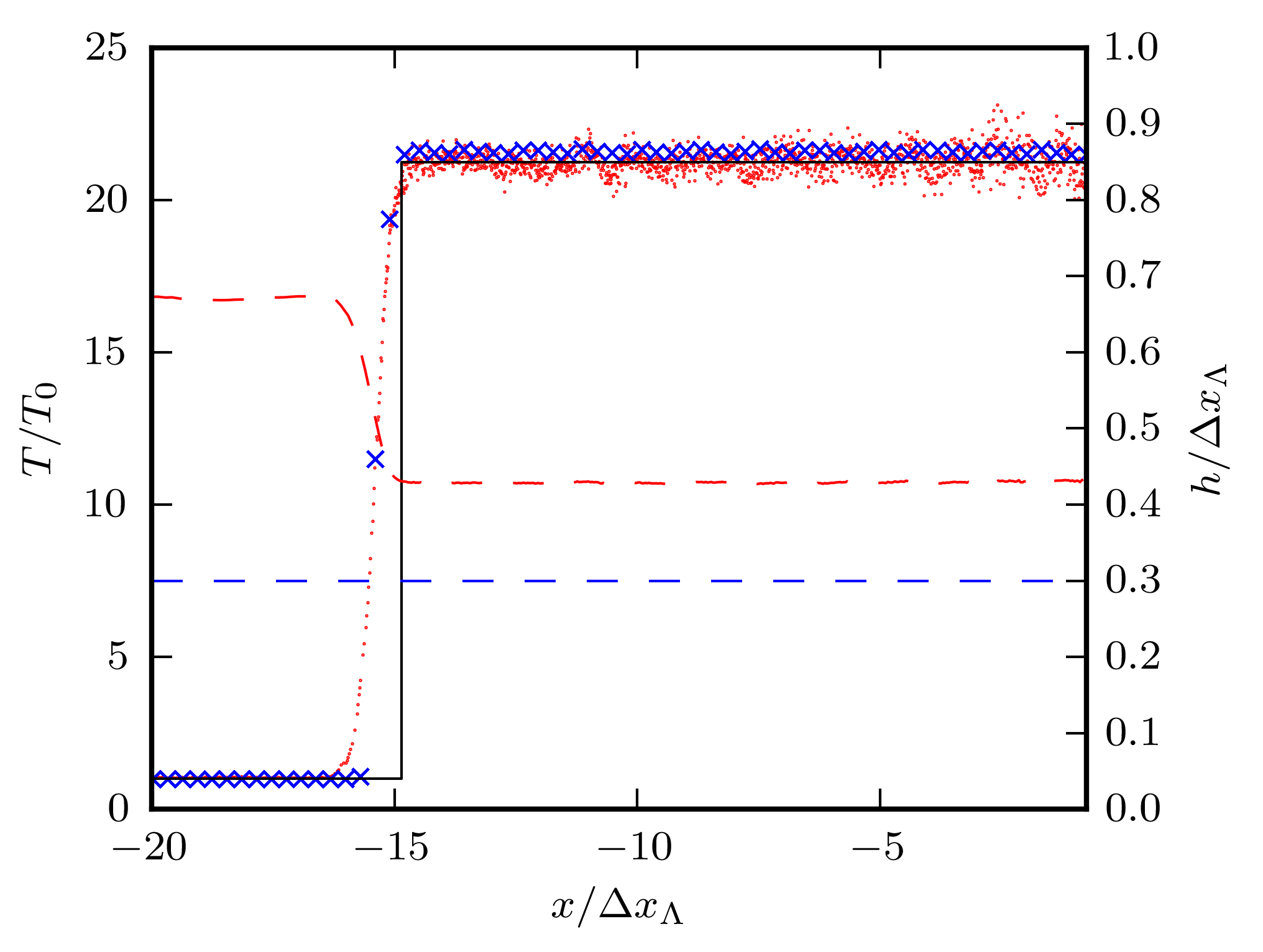}
\includegraphics[width=\columnwidth]{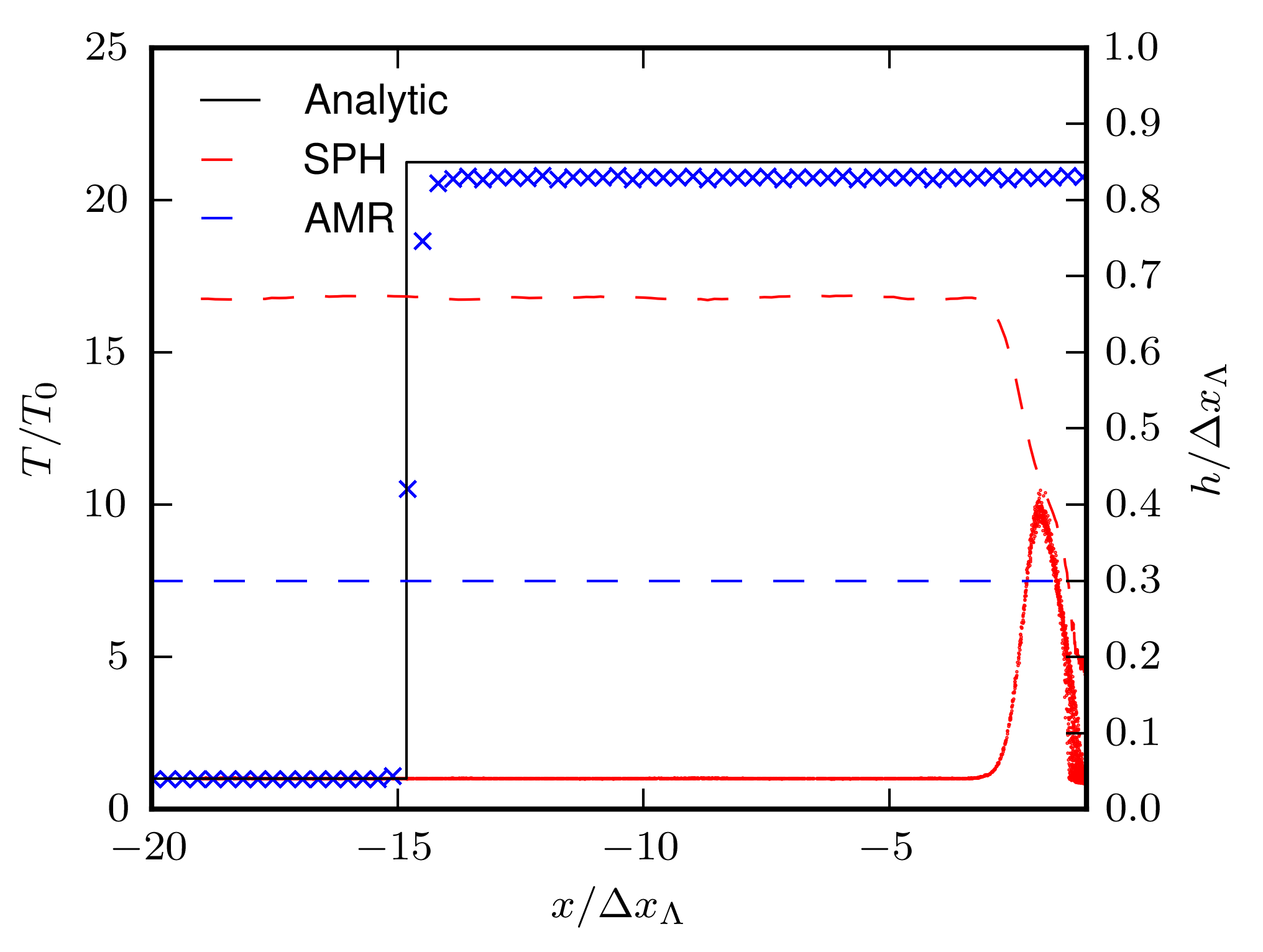}
\caption{As for Fig. \ref{fig:lowMach} but for an $\mathcal{M}=6.04$ shock. {\it Upper panel} shows the case without cooling, with a higher post-shock temperature than Fig. \ref{fig:lowMach}. 
\emph{Lower panel}, the case with cooling. Here the SPH
  particles shock over several smoothing lengths, allowing them time to
cool. Unfortunately, this means they never reach the higher temperature where cooling vanishes and their temperatures decline to the pre-shock value, forming a
cold dense region similar that in Fig. \ref{fig:lowMach}. The shift of the position of the shock front is due to the conservation of mass; cooling allows the post-shock gas to be compressed down to a small region around $x=0$. We note here that \flash\  adequately captures the post-shock temperature even when cooling is included.} 
\label{fig:highMach}
\end{figure}

First let us consider the case of cooling for the $\mathcal{M}=4.7$ shock. This should result
in a gas temperature of less than $(T_1 + T_0)/2$, i.e. we are
on the left side of the cooling spike. The initial collision of gas can
result in higher temperatures and follows an evolution for which we
have no analytic solution, before settling down to our stationary
case. 

In Fig \ref{fig:lowMach} (lower panel) we see the \flash\ and \gadget\ representations of
these shocks. Both codes reach a maximum temperature which is lower than that of the
similarity solution. In SPH we attribute this to \lq pre-shocking\rq\ , i.e particles will shock in several stages and cool as they are being shocked. In \flash\ we attribute this to the cooling operation being applied after the hydrodynamics in a time step, such that we do not record the post-shock temperature. Neither \gadget\ nor \flash\ has the resolution to reproduce the cooling tail particularly well here, although the final cold state is achieved in both cases.

For our second cooling test we look at a more extreme case, $\mathcal{M}=6.04$. This shocks the gas up to a temperature $T>T_1$ from
which it cannot cool, hence the analytic solution is the same as for a
shock without cooling. In Fig.~\ref{fig:highMach} we show the left hand
side of the shocked regions. Here the \flash\ simulation reproduces the analytical
result very well, but the \gadget\ simulation suffers from much more severe numerical overcooling through the pre-shock region, which prevents the gas from reaching the temperature from which it is
unable to cool (due to our choice of $\Lambda(T)$). As a result we see pile-up of high density cold gas around $x=0$, and the
shocked region is left far behind that of the \flash\ run\footnote{Note that if cooling
is prevented, the shock speed will be much higher relative to the rest frame. This is easily
understood in terms of conservation of mass, the gas is shocked to a
lower density and a much larger region is required to contain it.}. As a result of this overcooling the SPH simulation fails to form any hot gas at all. We note, however, that this is a general problem and not specific to either \gadget\, or SPH. 

\subsection{Convergence study for \gadget\ results}
\label{sec:Convergence}
\begin{figure}
\centering
\includegraphics[width=\columnwidth]{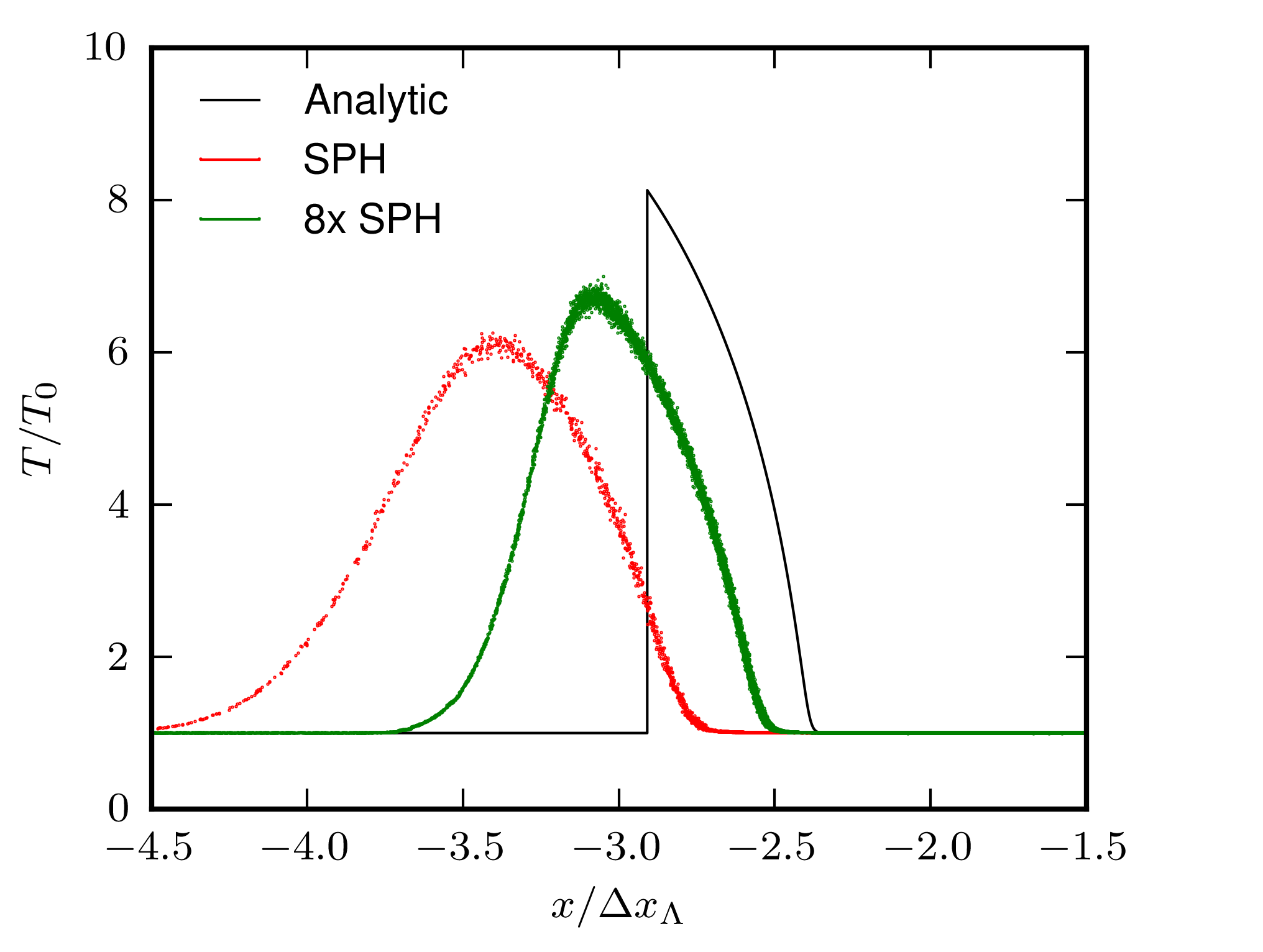}
\caption{As for the lower panel of Fig.~\ref{fig:lowMach}, but for two different SPH particle resolutions:{\it red points} are the SPH particles as for the lower panel of Fig. \ref{fig:lowMach}, whereas {\it green points} are for a $2 \times$ better resolution run (i.e. a factor of $8$ times more particles). The lower resolution run reproduces the temperature peak to within 25\%, for the higher it is around 15\%. When the simulation is close to
convergence, we would expect $\Delta T \propto \Delta u \propto h$, the
smoothing length, i.e to get within $1\%$ of the temperature we would
need a factor of $\sim 10^4$ more particles (compared to the lower resolution run). At higher resolutions the offset between the exact and simulated shock fronts is also reduced.
}
\label{fig:sphconvergence}
\end{figure}

As it stands, we can be confident that the results we have just given for SPH are not converged, as they have failed to reach our stable analytic solution. The problem we have attempted to solve involves no elements that an SPH code would not be expected to handle in the limit where the SPH resolution length $h \to 0$, and using a good prescription for artificial viscosity. The outstanding question here is thus only one of how much resolution is required; to this end we re-ran the $\mathcal{M}=4.7$ simulation with a factor of $8$ increase in the particle count\footnote{One can, of course, successively reduce the width of the shock tube in a 3d simulation to achieve the same scaling as a 1d one, i.e. $h \propto N_{\rm SPH}^{-1}$. In an astrophysical simulation, however, this option is usually unavailable as the shock will be embedded in an environment which needs to be simulated in full 3d.} (from $\approx 80,000$ to $\approx 660,000$). Let us first, however, make some general remarks about the problem.

Given the maximum temperature, $T_s$, of the radiating shock we can
estimate the error $\Delta T$ of the SPH maximum temperature by
estimating the radiative cooling over the shock (the physical shock is
non-adiabatic and so occurs on timescales many orders of magnitude
shorter than that of the cooling, as discussed in the Introduction). Assuming the width of the SPH shock
is $\sim h$, the temperature difference will be given, to first order,  by
\begin{equation}
\frac{\Delta T}{T_0}  \sim \frac{h}{v_0 k_B T_0} \cdot   m_p \left| \dot{u}_\Lambda  (T_s) \right| \sim  \frac{h}{\Delta x_\Lambda}\,,
\end{equation} 
where we have assumed that all the mechanical energy has been converted
into thermal energy, and that $T_s \gg T_0$. For larger smoothing
lengths we expect $\Delta T$ to become sub-linear in
the smoothing length, since the cooling is weaker at lower temperatures
(assuming we are on the left hand side of the \lq cooling spike\rq).

If we apply this argument to Figure \ref{fig:sphconvergence} we see
that increasing the number of particles by a factor of $\sim 8$
(i.e. reducing $h$ by a factor of $2$) reduces the temperature
error by a factor of $\sim 1.5$, suggesting that we are not quite
in the linear regime. To reach a temperature within $1\%$ of the analytic
temperature would seem to require increasing the particle count by
a factor of $\sim 10^4$. Although this is (barely) possible for this particular case, such resolution is not feasible in cosmological calculations. Most shocks in cosmological simulations will occur at lower resolution than we have used in this test case. Therefore we should seek an alternative solution involving a switch to prevent cooling during the shock process. We intend to explore such a switch in a further paper.

\section{A resolution criterion for radiative shocks}
\label{sec:resCrit}
Having established the difficulty of modelling some shock problems with radiative cooling we now wish to obtain a criterion against which we can judge simulations. Such a tool will allow us to identify those simulations where resolution is not a problem and those where more care is required. In the following section we will discuss the effects of resolution in quite a general way before deriving a metric from a simple model problem. We will frame our discussion in terms of SPH, however there will be analogous arguments for mesh codes.

Let us take a general case of a numerical simulation of a radiative shock. We assume we have pre-shock gas with velocity, specific internal energy and density $v,u, \rho$ which passes through a shock and comes to rest ($v$ is the velocity of the incoming gas with respect to that of the post-shock gas). First we note that the SPH shock has a width which is some multiple of the smoothing length $h \propto (m_{\rm SPH}/\rho)^{1/3}$ (for a mesh this would be the width of a cell), where the numerical factor will include some dependence on the artificial viscosity prescription. The change in thermal energy will be $\Delta u \propto v^2$ by energy conservation, and thus we can define a rate of \lq shock-heating\rq,

\begin{eqnarray}
\left. \dot{u} \right|_{shock} &=& \Delta u / \Delta t \\
& = & k v^3 (\frac{m_{\rm SPH}}{4 \pi \rho /3 })^{-1/3}\,,
\end{eqnarray}
where $k$ is some constant depending upon the details of the SPH scheme used (e.g neighbour counts). This heating rate is entirely numerical, as can be seen by the presence of the SPH particle mass $m_{\rm SPH}$: in the continuum approximation of the underlying fluid equations the shock heats the gas instantaneously, hence the heating rate is singular. As we reduce the particle mass, $h$ decreases and the numerical rate at which particles are heated over the shock front increases.

By taking the ratio of the physical rate of gas cooling to the numerical rate at which the gas is shock heated (which, ideally, we would wish to be almost infinite) we can analyse the effects of shock resolution. Only if the absolute value of this dimensionless ratio is small do we expect the shock heating to overwhelm the cooling, i.e. we require
\begin{equation}
\label{eq:generalResCrit}
|\dot{u}_\Lambda| \frac{1}{c^3\,{\cal M}^{3}} \left( \frac{m_{\rm SPH}}{\rho} \right)^{1/3} < \eta_{\rm SPH} \,,
\end{equation}

for the numerical solution to achieve close to the correct post-shock temperature, where $\eta$ is a dimensionless parameter. Here we have written the velocity of the incoming gas as $v={\cal M}c$ in terms of the Mach number and the 
upstream sound speed, $c\equiv c_0$. The equivalent for a mesh code can be written with the side length $h$ of a cubic mesh cell written in terms of the mass enclosed, $m_{\rm AMR} = \rho h^3$, to give
\begin{equation}
|\dot{u}_\Lambda| \frac{1}{c^3\,{\cal M}^{3}} \left( \frac{m_{\rm AMR}}{\rho} \right)^{1/3} < \eta_{\rm AMR} \,.
\end{equation}
In the subsequent section we attempt to determine a reasonable value of $\eta$ which we can use to evaluate other simulations.

\subsection{Heaviside cooling function}
To achieve an extremely simple radiative shock we set up a wall shock (see section \ref{sec:RadCoolingSetup}) with a low Mach number $\mathcal{M}=2$ and the piecewise cooling function
\begin{equation}
\dot{u}_\Lambda = \left( \frac{u_0^{3/2}}{\Delta x_0} \right) \cdot \left\{ \begin{array}{cc} 
0 , & u \leq u_0  \\
-\tilde{\Lambda} \left( \frac{\rho}{\rho_0} \right) ,&  u_0 < u  \end{array} \right.
\label{eq:heavi}
\end{equation}
where $\Delta x_0 \equiv (m_{\rm SPH}/\rho_0)^{1/3}$ is the initial interparticle spacing, $\tilde{\Lambda} > 0$ a dimensionless cooling parameter, $u_0$ the initial specific internal energy and as in Section~\ref{sec:RadCoolingSetup} we use $\gamma=5/3$. In the SPH simulation the particles are initially arranged on a cubic lattice of dimensions 1024x8x8 in units of $\Delta x_0$ (1024 referring to the long $x$ direction). The simulations were all performed with periodic boundary conditions. Usually a cooling function would be independent of the interparticle spacing, however we chose to re-use our initial conditions whilst adjusting the dimensionless constant $\tilde{\Lambda}$, and in this way scale the LHS of Eq.~\ref{eq:generalResCrit}. This is equivalent to using a fixed cooling function but adjusting the interparticle spacing.
\begin{figure}
\centering
\includegraphics[width=\columnwidth]{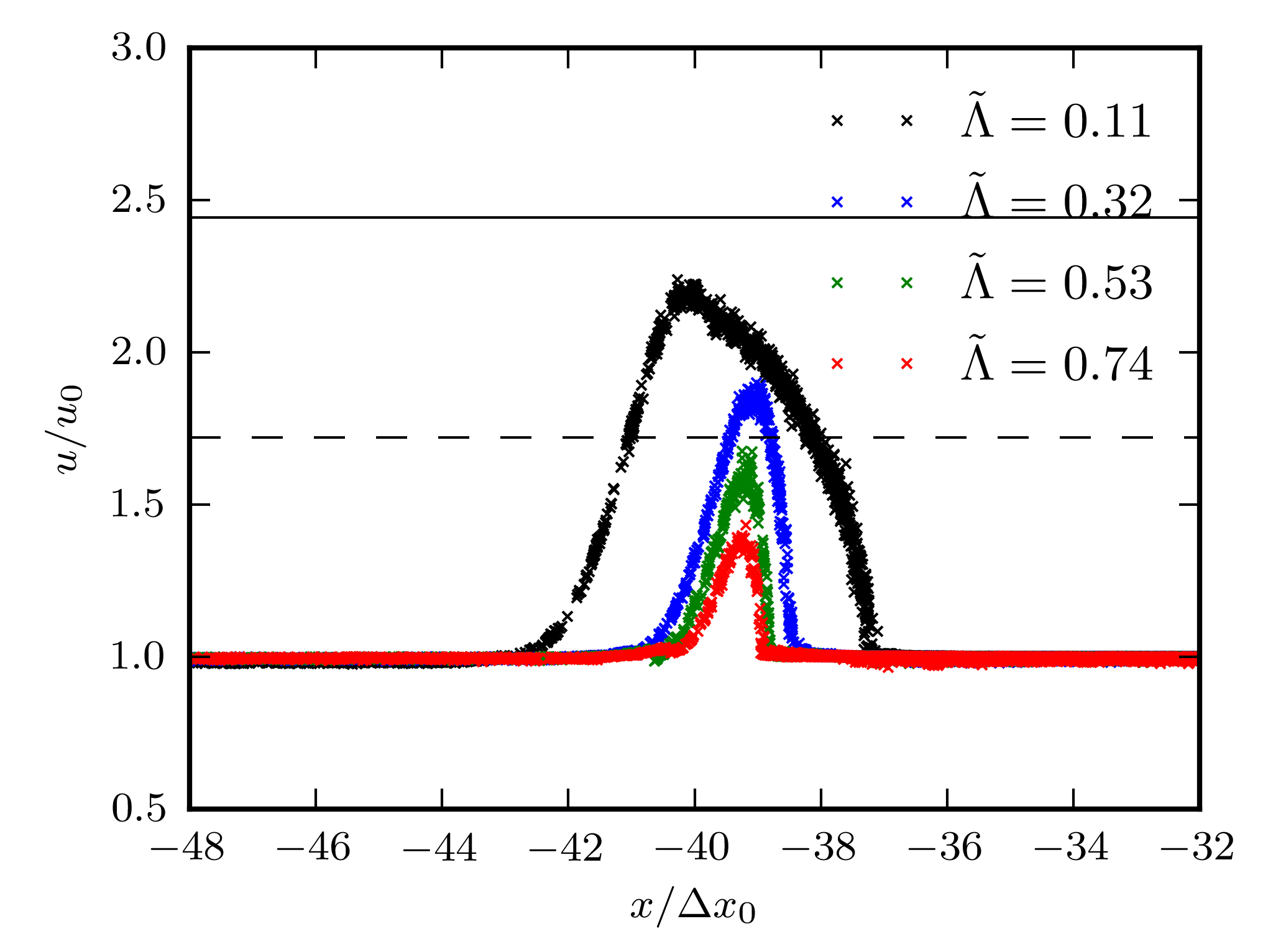}
\caption{Specific thermal energy vs. position for radiative shocks with a Heaviside cooling function, Eq.~(\ref{eq:heavi}). \emph{Black, blue, green, red crosses} are for SPH simulations with dimensionless cooling rates of $\tilde{\Lambda}$=0.11, 0.32, 0.53, 0.74, respectively; positions are quoted in units of the initial (pre-shock) particle spacing $\Delta x_0$, thermal energies in units of the initial thermal energy $u_0$. \emph{Solid line} indicates the analytic instantaneous post shock thermal energy $u_s/u_0 = 2.44$. \emph{Dashed line} indicates mid-point energy between the initial and final thermal energy $\frac{1}{2}(u_s+u_0)/u_0$. When the cooling rate is low ($\Lambda=0.11$, black crosses), the numerical overcooling is small and the simulation gets close to the right post-shock temperature. Increasing the cooling rate degrades the accuracy of the numerical result. We use this to set a maximum cooling rate that the simulation can tolerate, for example by requiring that the simulated post-shock temperature be larger than half the correct result (horizontal dashed line).}
\label{fig:mach2}
\end{figure}

We now make a couple of observations. Firstly, we note that we can calculate the instantaneous post-shock state using the equations derived in Appendix \ref{sec:Piecewise}, to find $\rho_s/\rho_0 =2.52$, ; $u_s/u_0 = 2.44.$ We note that this ratio is independent of the cooling parameter $\tilde{\Lambda}$. Increasing $\tilde{\Lambda}$ in the simulations, however, we expect the maximum post-shock temperature to fall as thermal energy is radiated away over the numerically broadened shock\footnote{One might have expected the post shock thermal energy ratio for a mach 2 shock to be precisely $2\gamma (\gamma-1) + 1=3.22$ as in the case without cooling, however the immediate post-shock region is still in motion w.r.t the final cold post-shock gas (hence  $\tilde{\Lambda} = 0$ is a special case).}.

In Fig.~\ref{fig:mach2} we see the results of these simulations plotted at a time of $t=141 u_0^{-1/2} \Delta x_0$. We note that the particle distribution in the pre-shock region has also diverged from a lattice arrangement (if it were still a lattice the particles would appear at multiples of $\Delta x_0$) into something more glass-like. This is to be expected as the SPH equations of motion favour a large distance to the nearest neighbour for a given density, which can be acheived with a close-packed or glass-like arrangement. The position, velocity and Mach number of the shock at late times are independent of the cooling function, for fixed $u_0$ (provided the cooling function restores the thermal energy of the gas to $u_0$) as is shown in Appendix \ref{sec:Piecewise}.

With a low cooling parameter ($\tilde{\Lambda}=0.11$, black crosses) we see that the post-shock thermal energy reaches near the theoretical value, whilst with a high cooling parameter ($\tilde{\Lambda}=0.74$, red crosses) we see that the simulation produces almost no hot gas. We take the mid-point of the thermal energies as a minimum value the code should reach to give at least approximately the correct answer. From Fig.~\ref{fig:mach2} this corresponds to a cooling parameter of $\tilde{\Lambda} \approx 0.4$. Substituting this maximum value into Eq.~(\ref{eq:generalResCrit}) allows us to evaluate the parameter $\eta$ as

\begin{eqnarray}
\eta &=& \tilde{\Lambda} u_0^{3/2}  \frac{1}{c^3\,{\cal M}^{3}} \left( \frac{\rho}{\rho_0} \right)^{2/3} \\
&=& \tilde{\Lambda} \left( \gamma \left(\gamma-1 \right)\right)^{-3/2} \mathcal{M}^{-3} \left( \frac{\rho}{\rho_0} \right)^{2/3} \\
&\approx & 0.08\,,
\end{eqnarray}
(where we have used the post-shock density $\rho=\rho_s = 2.52 \rho_0$), or, using Eq.~(\ref{eq:generalResCrit}),
\begin{equation} \label{eq:resCrit}
|\dot{u}_\Lambda| \frac{1}{c^3\,{\cal M}^{3}} \left( \frac{m_{\rm SPH}}{\rho} \right)^{1/3} < 0.08\,.
\label{eq:sphref}
\end{equation}
This is the value of $\eta$ we will use throughout the remainder of this paper.

A similar analysis with \flash\ yields an only slightly weaker criterion,
\begin{equation} \label{eq:amrCrit}
|\dot{u}_\Lambda| \frac{1}{c^3\,{\cal M}^{3}}\left( \frac{m_{\rm AMR}}{\rho} \right)^{1/3} < 0.09\,,
\label{eq:ref}
\end{equation}
where $m_{\rm AMR}$ refers to the mass contained in a mesh cell (since the mass in cells varies we have taken $m_{\rm AMR}$ to be the value in the cell immediately to the right of the shock, for consistency with the evaluation of $\rho$).

It is worth discussing the differences between a grid and an SPH scheme when the adaptive capabilities are utilised. SPH has a resolution (smoothing) length which refines in areas of high density as $\rho^{-1/3}$. AMR on the other hand can have much more general refinement criteria, for example allowing higher resolution to be applied on features which need not be dense (e.g. shocks). As such AMR has something of an advantage when it comes to shocks, as almost all refinement schemes will refine over discontinuous variable to the maximum level, and hence there is no need to impose the refinement criterion Eq.~(\ref{eq:ref}). Of course it is possible that a region of space in the simulation is already maximally refined, yet even so fails to satisfy the criterion Eq.~(\ref{eq:ref}). We can then interpret this as  a test of how well the finite resolution of an AMR simulation represents the physics 
in the problem.

To refine a simulation in a given volume $V$ of a 2 dimensional structure (such as a shock) to scale $h$ in SPH requires $N_{\rm SPH} \propto h^{-3}$ particles, whilst in AMR one would only need $ N_{\rm cell} \propto h^{-2}$ cells (we note that limitations on the refinement level between cells does not in general alter this scaling relation).

This can be contrasted with a sheet like structure in a vacuum (e.g. a gravitationally collapsed disk of thickness $\ll h$), which will only require $N_{\rm SPH} \propto h^{-2}$ particles, the same relation as AMR\footnote{As such SPH could be viewed as a refinement scheme optimised for gravitationally collapsed structures.}.

Note that we have been concentrating on how well the simulations reproduce {\em shocks} in the presence of cooling. In practise we would also like the code to correctly reproduce the {\em cooling tail}, {\em i.e.} the cooling of the gas once it has passed through the shock (the right-hand side of Fig.~\ref{fig:mach2}). Clearly here we would like to resolve the cooling length from Eq.~(\ref{eq:coolinglength}), by requiring that $\Delta x_\Lambda \gtsima h$  in SPH (or the cell size in mesh codes).

In the subsequent section we will apply the resolution criterion we derived to estimate in which areas of cosmological simulations 
numerical overcooling could be problematic.

\section{Effects of resolution on galaxy formation}
\subsection{Galaxy formation simulations}

\begin{figure}
\centering
\includegraphics[width=\columnwidth]{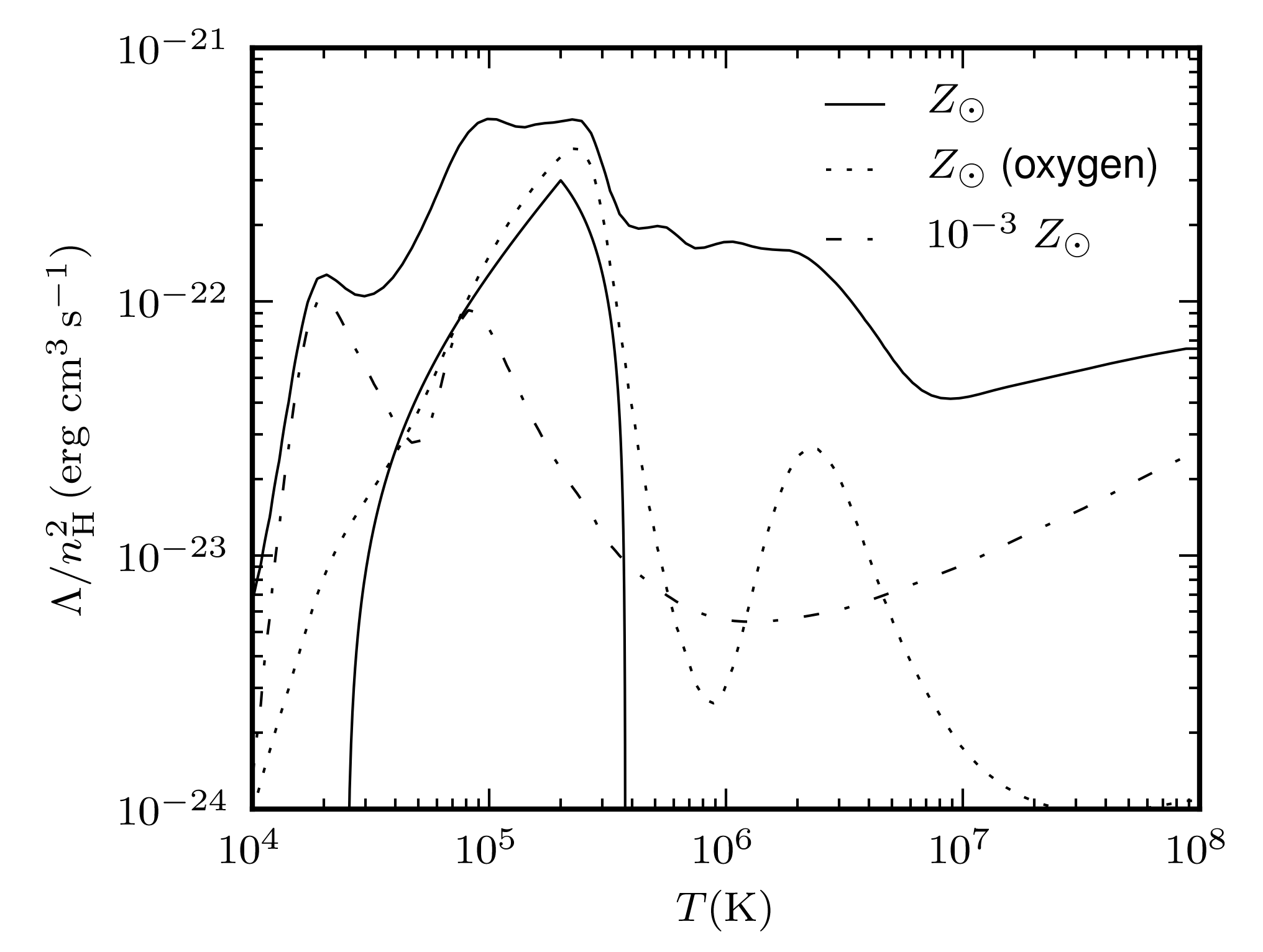}
\caption{Cooling functions used in the GIMIC simulations at redshift 0. \emph{Upper solid} and \emph{dot-dashed lines} represent astrophysical cooling functions for a plasma with  metallicity $[Z/Z_\odot]=0,-3$ (where square brackets denote the base-$10$ logarithm) respectively in the presence of an ionizing background \citep{Wiersma_Schaye_and_Smith_09}. \emph{Lower solid line} shows a cooling spike such as in section \ref{sec:RadCoolingSetup}, for comparison,  on the same logarithmic scale. \emph{Dotted line} shows cooling due to oxygen only, again assuming a solar abundance.}
\label{fig:gimiccool}
\end{figure}

\begin{figure}
\centering
\includegraphics[width=\columnwidth]{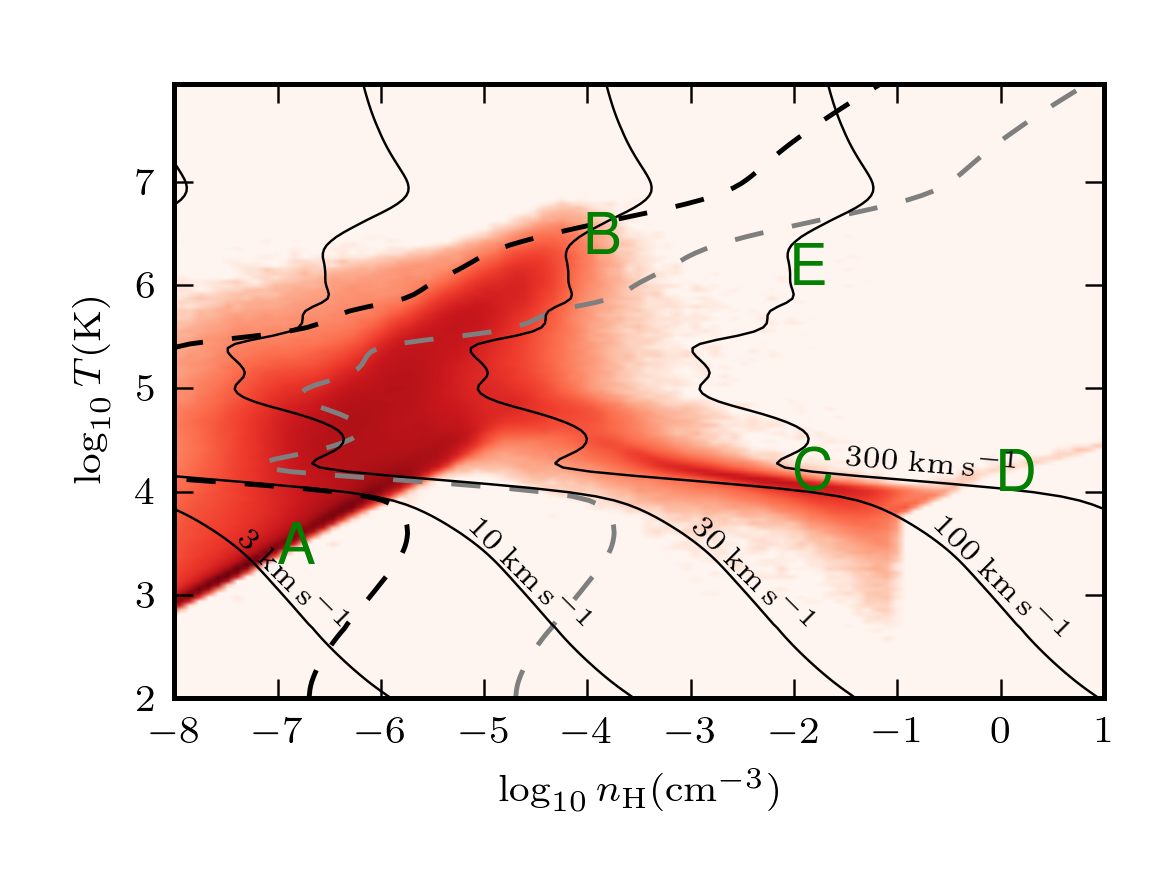}
\caption{Resolution requirements for correctly representing shocks in different regions of a temperature-density diagram. \emph{Solid contours} are labelled with the minimum values of the shock speed, $v=c\,{\cal M}$, obtained from Eq.~(\ref{eq:sphref}), that avoids excessive overcooling in the shock precursor, for simulations using an SPH particle mass of  $m_{\rm SPH}=10^6 M_\odot$. The radiative cooling rate adopted is that of a plasma with solar abundances of elements, $[Z/Z_\odot]=0$, as shown in  Fig.~\ref{fig:gimiccool}. The overlaid \emph{red shaded region} is the phase density in $(T,n_\mathrm{H})$ space of SPH particles in a cosmological feedback simulation (see text). \emph{Bold letters} refer to example environments described in Table~\ref{tab:Environments}. \emph{Heavy black} and \emph{grey dashed lines} refer to $t_{cool}=t_{dyn}$ and $t_{cool}=0.1t_{dyn}$ respectively. See text for discussion.}
\label{fig:velocity_resolution}
\end{figure}

In this section we apply the resolution criterion of Eq.~(\ref{eq:sphref}) to different regions of temperature and density in the the {\sc GIMIC} SPH simulation of \citet{Crain_et_al_09}.  First we will plot the distribution of gas in temperature-density space and identify some environments of interest. We will then discuss the radiative shock resolution in this parameter space, but also explore mitigating factors which may allow us to have confidence in simulations even when they fail to accurately resolve the shocks. 

The {\sc GIMIC} simulations are zoomed re-simulations of nearly spherical regions picked from the Millennium simulation \citep{Springel_et_al_05}, including gas dynamics. Each sphere has a radius of $18 h^{-1}$~Mpc, and the SPH particle mass is $m_{\rm SPH}\approx 10^6 h^{-1}M_\odot$. Radiative cooling in the simulation includes line cooling of eleven elements, Compton cooling with the CMB and thermal bremstrahlung in the presence of a uniform but evolving ionising background, as described in \citealp{Wiersma_Schaye_and_Smith_09} (see Fig.~\ref{fig:gimiccool}). The background cosmology, as for the Millennium simulation, is $\Omega_{\rm m}= 0.25$, $\Omega_\Lambda=0.75$, $\Omega_{\rm b}=0.045$, $n_s = 1$, $\sigma_8 = 0.9$, $H_0=100 \; h \; {\rm km \; {\rm s}^{-1} \; {\rm Mpc}^{-1}}$, $h=0.73$. The enrichment of gas by nucleosynthesis in stars is described in \cite{Wiersma2009}. Photo-heating, radiative and adiabatic cooling, shocks induced by galactic winds and due to accretion, result in gas occurring over a wide range of densities and temperatures, illustrated in Fig.~\ref{fig:velocity_resolution}.  Five points A-E in this $T-\rho$ space correspond to physical states where we want to investigate to what extent the simulation properly resolves radiative shocks (see also Table~\ref{tab:Environments}). For a general discussion of these diagrams see \citet{Wiersma_et_al_10}. The simulation code described here was also used in the {\sc OWLS} project \citep{Schaye2010}.

Point A is a typical IGM point outside virialised halos, at low density and temperature. Here we see a very well defined mild upward slope of temperature with $\rho$ of the post reionization gas. This gas is cooling due to adiabatic expansion of the universe and is being photo-heated by
the UV-background . For a recombination coefficient $\propto T^{-0.7}$ this will at late times result in a temperature-density relation of $T \sim \rho^{1/1.7}$ (\citealp{Hui_Gnedin_97,Theuns_et_al_98}).

Point B corresponds to gas heated in an accretion shock, falling into a galactic halo, or shocked by a galactic wind. Mechanical energy has been converted into thermal energy, and the density will jump by a factor up to $\sim 4$. When this gas cools, it will form the warm gas of point C which may condense to fuel star formation in a central  galaxy \citep{White_and_Rees_78}. 

On the far right, $n_\mathrm{H} > 10^{-1} \; \mathrm{cm^{-3}}$, is a sharp vertical feature in the distribution of SPH particles. This boundary delineates cold halo gas from gas which undergoes star formation in the model used in {\sc GIMIC}. The denser gas  represents a multi-phase interstellar medium, for which the imposed pressure-density relation in {\sc GIMIC} is $p\propto n_{\rm H}^{4/3}$, known as an effective equation of state for the ISM. Such a state is intended to mimic the physical pressure response in dense gas undergoing star formation (point D), compressing this gas results in significant star formation with associated feedback (see \cite{Schaye_and_Dalla_Vecchia_08} for motivation and details). The SPH density then represents a volume average density of star-forming gas, whereas the physical ISM lies in approximate pressure equilibrium, but with a hot and cold phase and corresponding variation in densities. In particular the simulation does not allow this gas to cool radiatively. Finally, point E represents the domain of type II SNe that ignite in the hot ($10^6$K) sparse phase of the ISM, generated by the activity of previous generations of SNe. We note that there is little gas marked in this phase as the cooling time is short.

Now let us consider this simulation in terms of its ability to resolve the radiative shocks 
that occur in these 5 regions. Equation (\ref{eq:resCrit}) suggests that a radiative shock of velocity $v$ will be resolved if it satisfies
\begin{equation}\label{eq:velCrit}
 v > \left( \frac{|\dot{u}_\Lambda |}{ 0.08} \right)^{1/3} \left( \frac{m_{\rm SPH}}{\rho} \right)^{1/9} \,,
\end{equation}
{\em i.e.} there is a minimum shock velocity which can be resolved. Shocks
below this velocity will tend to artificially radiate away their
energy because there will be cooling through the (artificially extended) shock region. Shocks above this velocity will heat up the gas on such a short timescale in the simulation that the cooling in the shock region will make little difference to the final result.

In Fig. \ref{fig:velocity_resolution} we plot contours of given $v$, the minimum shock velocity for which there is no significant overcooling in shocks. At each temperature and density a cooling rate is evaluated (using the cooling rate from \citealp{Wiersma_Schaye_and_Smith_09}, shown in Fig.~\ref{fig:gimiccool}, evaluated at solar metallicity, 
and in the absence of an ionising background), which is combined with a particle mass of $m_{\rm SPH}=10^6 M_\odot$, to derive a minimum shock velocity which can be resolved. Note that Eq. (\ref{eq:velCrit}) is very weakly dependent on particle mass, and thus changing mass resolution is a very ineffective way of shifting the contours. These contours represent the minimum velocity shock which can be resolved at each $T,\rho$. Any shocks at lower velocities will appear artificially colder due to resolution effects.

\begin{table}
  \begin{center}
    \begin{tabular}{|  l | c | c | c | c | c |}
\hline
	 & $n_H$ & $T$ & $\Lambda$ & $h$ & $v_{min} $ \\
	& (cm$^{-3}$) & (K) & (erg cm$^3$/s)  & (kpc) & (km/s) \\
    \hline
 A. IGM & $10^{-7}$ & $2000$ & $10^{-24}$  & $150$& $5$ \\ 
  B. Hot halo  & $10^{-4}$ &  $2 \times 10^6$ &  $10^{-22}$ & $15$ &  $100$\\
  C. Cold halo  & $10^{-2}$ &  $10^4$ &  $5 \times 10^{-24}$ & $3$ &  $200$\\
  D. ISM (AGN) & $10^0$ & $10^4 $ & $5 \times 10^{-24}$  & $0.7$&  $400$\\
  E. ISM (SNe) & $10^{-2}$ & $10^6$ & $10^{-22}$ & $3$&  $400$\\
    \hline
    \end{tabular}
  \end{center}
  \caption{Astrophysical shock environments identified in Fig.~\ref{fig:velocity_resolution}.}
\label{tab:Environments}
\end{table}

A key point, however, is that even if we fail to resolve the radiative shock, the cooling of the gas in many cases may be inevitable anyway. Indeed, there can be situations where other processes are occurring on much longer timescales than the cooling, and for which having an incorrect thermal history of the gas is not a problem as far as the dynamics of the system is concerned\footnote{Note that even if dynamics is not affected, there may be other consequences of numerically underestimating the amount of hot gas, for example when calculating the spectrum of cooling radiation.}. Establishing a general criterion for these is not trivial, here we will simply compare to the locally estimated dynamical time $t_{dyn}\equiv (G\rho)^{-1/2}$ as indicative of the timescales for other processes. We assume that the simulation will cool adequately if $t_{dyn} \gg t_{cool}$ even in the case that radiative shocks are resolved poorly (we define $t_{cool} \equiv \left| \dot{u_\Lambda}\right| /u$). The heavy dashed black contour in Fig.~\ref{fig:velocity_resolution} represents the line where $t_{dyn}=t_{cool}$. All points to the left of this represent $t_{dyn}<t_{cool}$ so certainly we would wish to completely resolve any shocks here. We have also included in dashed grey a line where $0.1 t_{dyn}=t_{cool}$ to demonstrate a somewhat stronger limit. The necessity of resolving the thermal history to the right of this line, is questionable, because the gas cooling time is so small in any case. Of course these simulations assume ionisation equilibrium and
optically thin gas, so the cooling rates may have been overestimated.
In addition if one were to attempt to track the chemistry of the shocked gas, for example the formation and destruction of molecular hydrogen, then having a correct thermal history could still be important \citep{Abel_et_al_97}.

Now let us evaluate the resolution criterion of Eq.~(\ref{eq:velCrit}) for the
five diverse environments of Table~\ref{tab:Environments}. First we take point A, for radiative shocks in the IGM; here we can see that the low density and cooling rate combine to allow us to resolve all shocks above $5$~km~s$^{-1}$, almost certainly much exceeding our requirements.

For point B we have taken a value for gas heated by a virial shock to $2 \times 10^6$ K. The higher density and cooling rate here push up the minimum shock velocity we can resolve to around $100$~km~s$^{-1}$, comparable to the virial shock velocities $v_{200}$ themselves for halos of mass $\sim 10^{12}M_\odot$:
\begin{equation}
v_{200} = \left( 10 G H(z) M_{200} \right)^{1/3} ,
\label{eq:vvir}
\end{equation}
\citep{Mo_Mao_and_White_98},  where $G$ is the gravitational constant, $H(z)$ the Hubble parameter and $M_{\rm 200}$ is the virial mass of the halo. For $z\gtrsim 1$ we can approximate the Hubble parameter as $H(z) \approx H_0 \Omega_m^{1/2} (1+z)^{3/2}$ to see
\begin{eqnarray}
v_{200} &\approx & 201\,{\rm km}\,{\rm s}^{-1}\,\left({h\over 0.73}\right)^{1/3}\,\left({\Omega_m\over 0.25}\right)^{1/6} \times \nonumber \\
& & \left({1+z\over 3}\right)^{1/2}\,\left({M_{200}\over 10^{12}M_\odot}\right)^{1/3} , \\
\label{eq:tvir}
T_{200} &=& \frac{1}{3} \frac{\mu m_{\rm p}}{k_{\rm B}} v_{200}^2 \\
&\approx& 1.0\times 10^6\,{\rm K}\,\left({h\over 0.73}\right)^{2/3}\,\left({\Omega_m\over 0.25}\right)^{1/3} \times \nonumber \\ 
& &\left({1+z\over 3}\right) \left({M_{200}\over 10^{12}M_\odot}\right)^{2/3}\, ,
\end{eqnarray}
where $\Omega_m$ is the matter density in units of the critical density and $h= H_0 / 100 \rm \, km \, s^{-1} \, Mpc^{-1}$. For lower mass haloes the gas actually cools even faster and the shocks are more difficult to resolve, however the cooling may be so short as to save the situation. We explore this situation further in Section~\ref{sec:VirialShocks}. 

For point C we consider the warm gas within galactic disks. The minimum shock velocity which can be resolved close to the star formation threshold is higher than for point B, because the cooling rate is higher. However the cooling time is so much smaller than the dynamical time in the disk, so that we suspect that gas cooling will be inevitable in any case. This suggests that, although numerical overcooling is potentially severe here, it is unlikely to have any important effects on the evolution of the disc.

At a higher temperature than C we have point E, a fiducial point for a (type II) supernova blast wave shocking to hot ($\sim 10^6$K), rarefied ($n_{\rm H} \sim 10^{-2}$ cm$^{-3}$) ISM (irradiated and inflated by the massive progenitor star). We have a high minimum resolved shock velocity due to the fast cooling of this gas, making its simulation problematic. We expect the gas to form a cold, dense shell \citep{Cox_72}, and the lack of resolution to manifest itself primarily in an alteration of the onset of this phase. We discuss the implications for SN feedback further in Section~\ref{sec:supernovae}.

Finally, for point D we have considered the case of an AGN outflow shocking into a dense ISM of $n_H \sim 1$ cm$^{-3}$. The minimum resolved velocity is so high here that we can have little confidence in the simulated shock dynamics (excluding the most basic properties such as conservation of momentum). The gas is cooling fast compared to dynamical time scales, yet we have similar concerns to point E about the artificial suppression of feedback.

\subsection{Virial Shocks}
\begin{figure}
\centering
\includegraphics[width=\columnwidth]{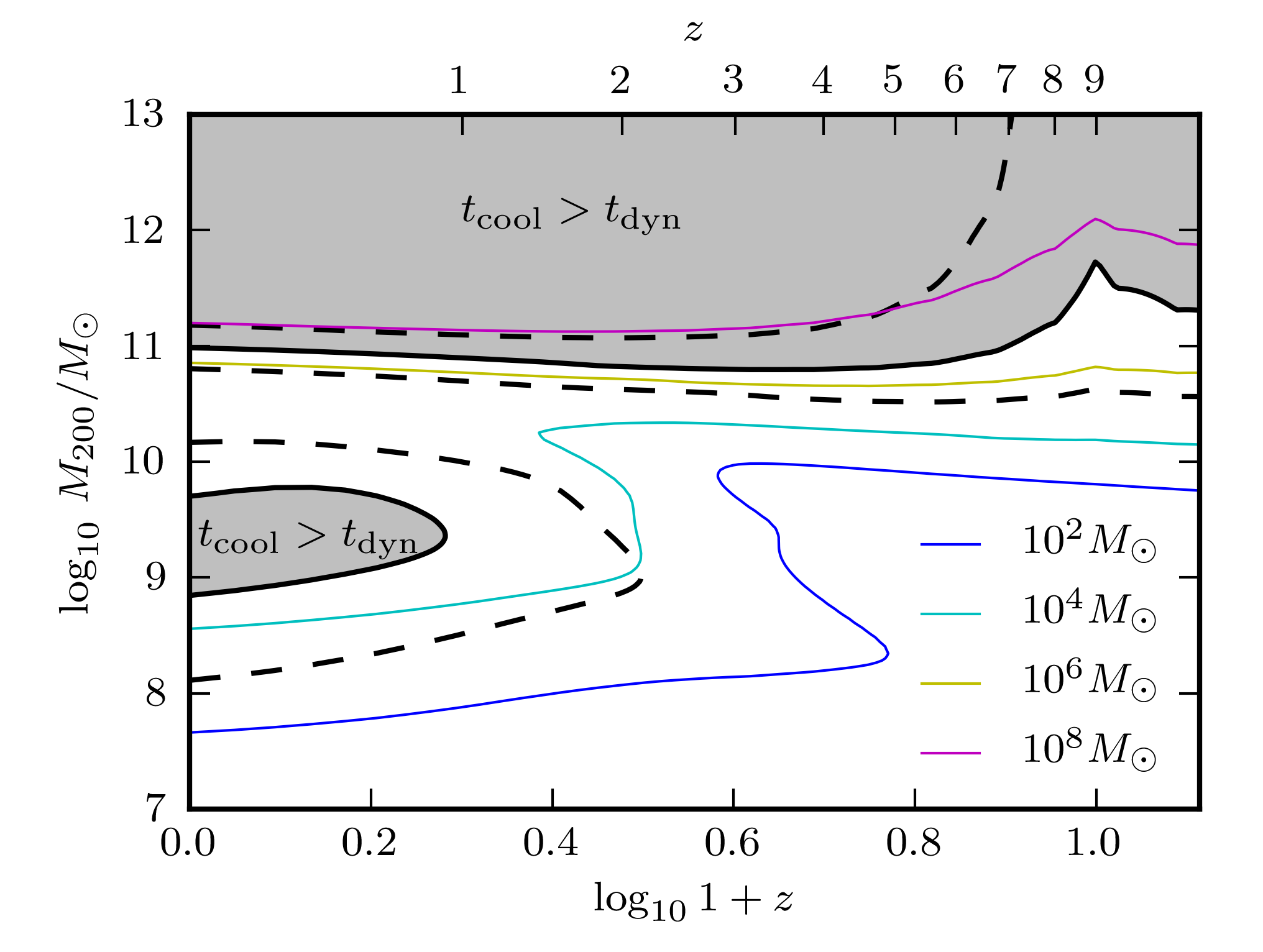}
\caption{Contours of maximum SPH particle mass $m_\mathrm{SPH}$ required to
  prevent numerical overcooling at a virial shock, for gas ($[Z/Z_\odot]=-3$) accreting onto haloes of different virial masses $M_{200}$ at a given redshift $z$. Coloured lines corresponding to $m_\mathrm{SPH}=10^8, 10^6, 10^4, 10^2 M_\odot$ limits are
  represented by the \emph{thin maroon, yellow, cyan} and \emph{blue lines} respectively. \emph{Black lines} compare cooling time to the dynamical time of the halo: 
  $t_{cool}=t_{dyn}$ (\emph{heavy solid line}), $t_{cool}=2 t_{dyn}$, $t_{cool}=\frac{1}{2} t_{dyn}$ (\emph{heavy dashed lines}).
The \emph{shaded grey} region denotes $t_{cool}>t_{dyn}$. Numerical overcooling due to lack of resolution in regions where  $t_{cool}\gtsima t_{dyn}$ will likely affect the dynamics of the accreting gas, hence SPH simulations would appear to need resolutions $\sim 10^6\,M_\odot$.}
\label{fig:virial_resolution}
\end{figure}

\begin{figure}
\centering
\includegraphics[width=\columnwidth]{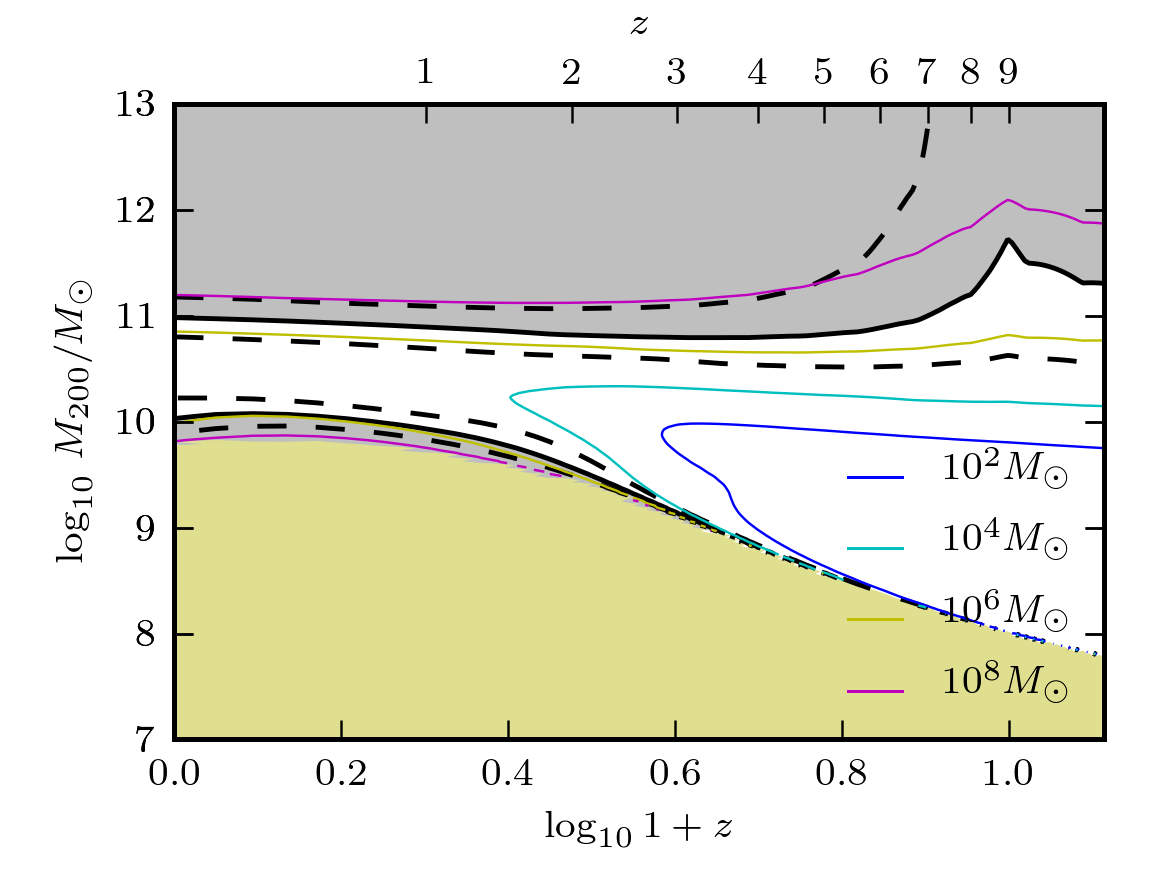}
\caption{As for Figure \ref{fig:virial_resolution}, this time including a uniform ionising background, see text for details. In the \emph{gold region} (lower left of the figure), the gas is being heated rather than cooled, so resolution of the shock is of lesser importance.}
\label{fig:virial_uvback}
\end{figure}

\label{sec:VirialShocks}
We now consider the effects of the resolution requirement Eq.~(\ref{eq:sphref}) on the discussion of cold accretion and virial shocks around haloes. Here we are following the ideas of spherical collapse set out in \citet{White_and_Rees_78}. The basic question here is what is the fate of gas as it accretes on to a halo, and gets shocked as it converts its mechanical energy into thermal energy. If the cooling time is short, then this hot phase will be a temporary one, however if the cooling time is long, then a hydrostatic hot halo of gas will form within the halo, the scaling relations for which can be found in {\em e.g.} \citet{Mo_Mao_and_White_98}.

The properties and stability of such spherical shocks have been studied by {\em e.g.}  \cite{Birnboim_and_Dekel_03}, depend on mass and redshift. More massive haloes have hotter shocks with longer cooling times. At a given mass
lower redshifts imply lower densities and hence slower cooling, and hence 
easier build-up of a hot halo. It must be recalled, however, that in this situation geometry will also play a role. If the gas accretion can achieve a configuration where it will penetrate farther into the halo ({\em e.g.} in filaments), it will shock at higher densities and generally have a shorter cooling time (the situation is complicated by the fact that the gas continues to accelerate, and so the shock will generally be hotter).

Here we first consider applying our resolution criteria to the spherical case. Assuming a spherical halo of mean density 
\begin{equation}
\bar{\rho}_{200} = 200 \left( \frac{3H^2}{8\pi G} \right) \, ,
\end{equation}
and virial mass $M_{200}$, we take the accreting gas to shock to the virial temperature\footnote{The infalling gas has actually twice this energy so if it shocks into the rest frame of the halo the temperature will be increased by a factor of 2, however we will ignore this for now.} $T_{200}$ and virial velocity $v_{200}$ given in Eqs.~(\ref{eq:vvir}) and (\ref{eq:tvir}), respectively. For the baryon density at the edge of the halo we use
\begin{equation}
\rho_{\rm b} = \frac{1}{3} \frac{\Omega_{\rm b}}{\Omega_{\rm m}} \rho_{200},
\end{equation}
where the factor $1/3$ is the ratio of edge to mean densities for an isothermal sphere of profile $\rho = \rho_0 (r/r_0)^{-2}$. We can then apply our shock resolution criteria in terms of the maximum mass of SPH particles that do not suffer from numerical overcooling in the shock,

\begin{equation}
m_{\rm SPH, max} =  \eta^3 |\dot{u}_{\Lambda(T_{200})}^{-3}|\, v_{200}^9\, \rho_{\rm b}\,,
\label{eq:msph}
\end{equation}
where our convergence tests suggest that $\eta\approx 0.08$. 

Equation~(\ref{eq:msph}) defines curves in a plot of virial mass $M_{200}$ versus redshift $z$, shown in Fig.~\ref{fig:virial_resolution} for a cooling rate appropriate for a plasma with solar abundance ratios but mean metallicity of $[Z/Z_\odot] =-3$ (we have chosen the lower metallicity as more representative of accreting gas that has yet to be enriched by several generations of star formation). In Fig.~\ref{fig:virial_uvback} we show the case where cooling is partly suppressed by the presence of a uniform ionising background (see \cite{Wiersma_Schaye_and_Smith_09} for details). Each thin coloured line represents the limiting particle mass required to prevent numerical overcooling in the corresponding spherical accretion shock. Clearly the resolution requirement is punitatively strict (smallest $m_{\rm SPH}$) for small haloes ($M_{200}\sim 10^{8-10}M_\odot$), especially at high redshifts ($z\sim 9$). Intuitively this can be understood because these masses correspond to virial temperatures near the peak of the cooling function, and at higher redshifts the mean baryon density (and thus collisional cooling rate) grows. In the presence of an ionising background cooling is suppressed in lower-mass haloes that cool mostly through hydrogen lines. At lower masses the ionising radiation has a very large effect, and gas will be {\em photo-heated} instead of cooling \citep{okamoto08}.

However, even though lack of resolution will lead to overcooling in some haloes, the cooling time in these haloes may be so short that the gas would cool quickly anyway.
The grey area in the figure indicates where the dynamical time in the halo is shorter than the cooling time: in this region we expect that numerical overcooling may prevent the formation of a hot halo. Conversely, in the white region, cooling is so fast that even if a hot halo were to form, it would quickly cool. The demarcation line between these scenarios follows closely the $\sim 10^6 M_\odot$ limiting SPH mass (yellow line). Simulations run with that resolution or better will be able to form hot spherical haloes in situations where we would expect such a hot halo to form. At lower resolution, simulations may artificially suppress the formation of a hot halo due to numerical overcooling in the accretion shock. 

Our considerations apply only at the virial radius. However, nearer the centre of halos we expect this conclusion to remain valid, as the cooling time diminishes faster than the dynamical time. The very high mass halos have $2 t_{dyn}<t_{cool}$ (heavy dashed black line) and we expect these to be in near hydrostatic equilibrium. As a result of these analyses we conclude that $10^6 M_\odot$ is a sensible upper limit for the gas particle mass in cosmological simulations intending to capture the evolution of proto-galactic haloes, although lower masses enable more accurate resolution of the thermal history of gas in lower mass haloes.

\subsection{Thermal feedback}
\label{sec:supernovae}
Thermal feedback refers to the mechanism whereby injection of thermal energy into the ISM causes adiabatic expansion of the gas and subsequent suppression of star formation due to the diminished density. The simplest model to envisage is perhaps that of a single supernova creating a hot, spherical, rarefied, blast wave. On larger scales, however, we expect to see analogous effects from star forming regions and AGN. In this section we intend to consider our results in terms of thermal feedback in SPH. We will review the basic physics of thermal feedback and its role in galaxy evolution. We will then discuss its implementation in SPH and derive some quantitative criteria for accurately resolving it. Finally, we will relate our observations to the feedback experiments in other work.

We begin by considering the problem of simulating a supernova blast wave. Here we are primarily concerned with the situation where we may artificially radiate away the thermal energy of the blast wave due to a lack of resolution. This would result in the premature transition from a thermally driven to a momentum driven phase.

A concise overview of the evolution of a supernova remnants can be found in \citet{Cox_72}. Essentially the blast wave will follow a Sedov-Taylor self-similar solution \citep{Sedov_59} until the shock temperature $T_s$ falls to a value where the radiative cooling exceeds the cooling via adiabatic expansion. A full calculation is beyond the scope of the present paper, however, we can get close just by dimensional considerations
\begin{eqnarray}
k_B T_s &=& \left( E_0^2 m_\mathrm{p}^3 n_\mathrm{H}^4 \Lambda^6 \right)^{1/11} \\
& \approx & k_B \cdot 4 \times 10^6 \mathrm{K}\, , 
\end{eqnarray}
where $E_0 = 10^{51}$ erg is the SNe energy, $n_\mathrm{H}=1.0$ cm$^{-3}$, $\Lambda=10^{-22}$ erg cm$^3$/s. The value of \citealt{Cox_72} is a factor 2 smaller, at $2.0 \times 10^6$K.

In a simulation we would like to resolve the transition in the supernova shock from being pressure driven to being momentum driven, which typically occurs for shock temperatures $T_s \sim 2.0 \times 10^6 $K; numerical overcooling may cause the shock to transition too early, hence underestimating the feedback effect of the explosion on star formation in the surroundings. Using the Sedov similarity solution for a 3 dimensional blast wave in a uniform cold medium of adiabatic index $\gamma=5/3$ and density $\rho_0$ (which we will shortly take to be the density of the ISM), we can then write the pressure and temperature just inside the shock wave in terms of the shock velocity $v_s$, as

\begin{eqnarray}
\rho_s &=& \frac{\gamma+1}{\gamma-1} \rho_0 \\
p_s &=& \frac{2}{\gamma+1} \rho_0 v_s^2 \\
k_B T_s &=& \bar{\mu} \frac{p_s}{\rho_s} =2 \frac{ \gamma -1}{(\gamma+1)^2} \bar{\mu}  v_s^2 ,
\end{eqnarray}
where $\bar{\mu}$ is the mean particle mass. Combining this with the resolution criterion of Eq.~(\ref{eq:resCrit}) we find that (excluding fairly pathological cooling functions) the cooling will be hardest to resolve at the lower temperatures, {\em i.e.} at $T_s = 2.0 \times 10^6 $~K.

Applying fiducial values for the ISM of $\bar{\mu} \approx 0.6\,m_p $ at $T_s = 2.0 \times 10^6 $~K yields a shock velocity, density and cooling rate of
\begin{eqnarray}
v_s &=& 380 \; {\rm km}\,{\rm s}^{-1} \\
\rho_s &\approx& 9 \times 10^{-24}\;{\rm g}\,{\rm cm}^{-3}\\
\left. \dot{u} \right|_\Lambda &\approx& -180 \; {\rm cm }^2\,{\rm s}^{-3}\,,
\end{eqnarray}
and the corresponding radius of the blast wave
\begin{eqnarray}
r_s &=& 1.15 \left( \frac{E}{\rho_0} \right)^{1/5} t^{2/5} \\
 &=& 1.15^{5/3} \left( \frac{E}{\rho_0} \right)^{1/3} \left( \frac{2}{5}\right) ^{2/3} v_s^{-2/3} \\
&\approx& 15 \; \mathrm{pc}\,,
\end{eqnarray}
where $E\sim 10^{51}{\rm erg}$ is the thermal energy injected by the explosion.
The corresponding limiting SPH particle mass that avoids numerical overcooling, evaluated from Eq.~(\ref{eq:sphref}), is 
\begin{eqnarray}
m_{\rm SPH} &=& 70 M_\odot 
\left( \frac{\rho_s}{9 \times 10^{-24} {\rm g}\,{\rm cm}^{-3}}\right)\nonumber \\
&\times &\left( \frac{v_s}{380 \; \mathrm{km}\,{\rm s}^{-1}}\right)^9 \left( \frac{\left| \dot{u_\Lambda}\right|}{180 \; {\rm cm^2}\,{\rm s}^{-3}}\right)^{-3}\,.
\end{eqnarray}

The small values of both the radius and the required minimal mass resolution imply that most cosmological simulations of galaxy formation are far from resolving individual SN explosions, however detailed simulations of high-$z$ dwarf galaxies do indeed already reach such extreme resolutions ({\em e.g.} \cite{Wise2008}). For a state of the art mass resolution for a cosmological simulation of say $10^5\,M_\odot$, a star particle really represent very many stars and hence also many SNe. Simply scaling-up $E_0$ to represent the many SNe that go off does not really help much, as for example the blast radius only scales $\propto E^{1/3}$. In reality different SNe will go off in different places, and once the density of the ISM is decreased due to one explosion, another explosion in the lower density gas will have a much larger effect, eventually resulting in a percolating hot phase.

However these small scales cannot yet be resolved in current cosmological simulations, hence they fail to follow the transition from pressure to momentum-driven SN shells. One can try to model the expected effects by simply heating a small number of neighbouring SPH particles. In this case our resolution study indicates that the reheating temperature must be sufficiently high for numerical overcooling not to affect the dynamics.  We can thus generalise the above calculation to find the minimum temperature for resolved thermal feedback for a given SPH particle mass.  To perform this calculation we will need to associate a shock velocity with a single particle, which we will take to be the Sedov shock velocity for a blast wave of the same mass as the SPH particle
\begin{eqnarray}
v_s &=& 1.15^{5/2} \sqrt{\frac{\pi}{3\gamma(\gamma-1)}} \cdot \frac{4}{5} c_s \\
&=& 1.1 c_s\,.
\end{eqnarray}
Combined with Eq.~(\ref{eq:velCrit}) we find
\begin{eqnarray}
u_\mathrm{fb} &\gtsima & \frac{\eta^{-2/3}}{1.1^2 \cdot \gamma (\gamma - 1) } \left( \frac{m_{\rm SPH}}{\rho} \right)^\frac{2}{9} (|\dot{u}_\Lambda |)^{2/3} \\
 & =& 4 \left( \frac{m_{\rm SPH}}{\rho} \right)^\frac{2}{9} (|\dot{u}_\Lambda |)^{2/3}\,, \label{eq:sneCrit}
\end{eqnarray}
the minimum thermal energy required to avoid numerical overcooling.

\citet{Dalla_Vecchia_and_Schaye_08} argue that for thermal feedback to be effective requires that the sound crossing time across an SPH particle, $t_s=h/c_s$, be smaller than the cooling time, $\tau_c=u/|\dot u_\Lambda|$. Using $\rho\,h^3\sim m_{\rm SPH}$, $t_s<\tau_c$ requires that
\begin{equation}
u_\mathrm{fb} \gtsima \left({m_{\rm SPH}\over\rho} \right)^{2/9}\, (|\dot{u}_\Lambda |)^{2/3}\,, \label{eq:sneCrit2}
\end{equation}
which is identical apart from a numerical factor to Eq.~(\ref{eq:sneCrit}). Our criterion is stronger as it takes into account that the code will in practise overestimate the cooling of the gas in shocks; the lower value simply requires there to be a shock.

\begin{figure}
\centering
\includegraphics[width=\columnwidth]{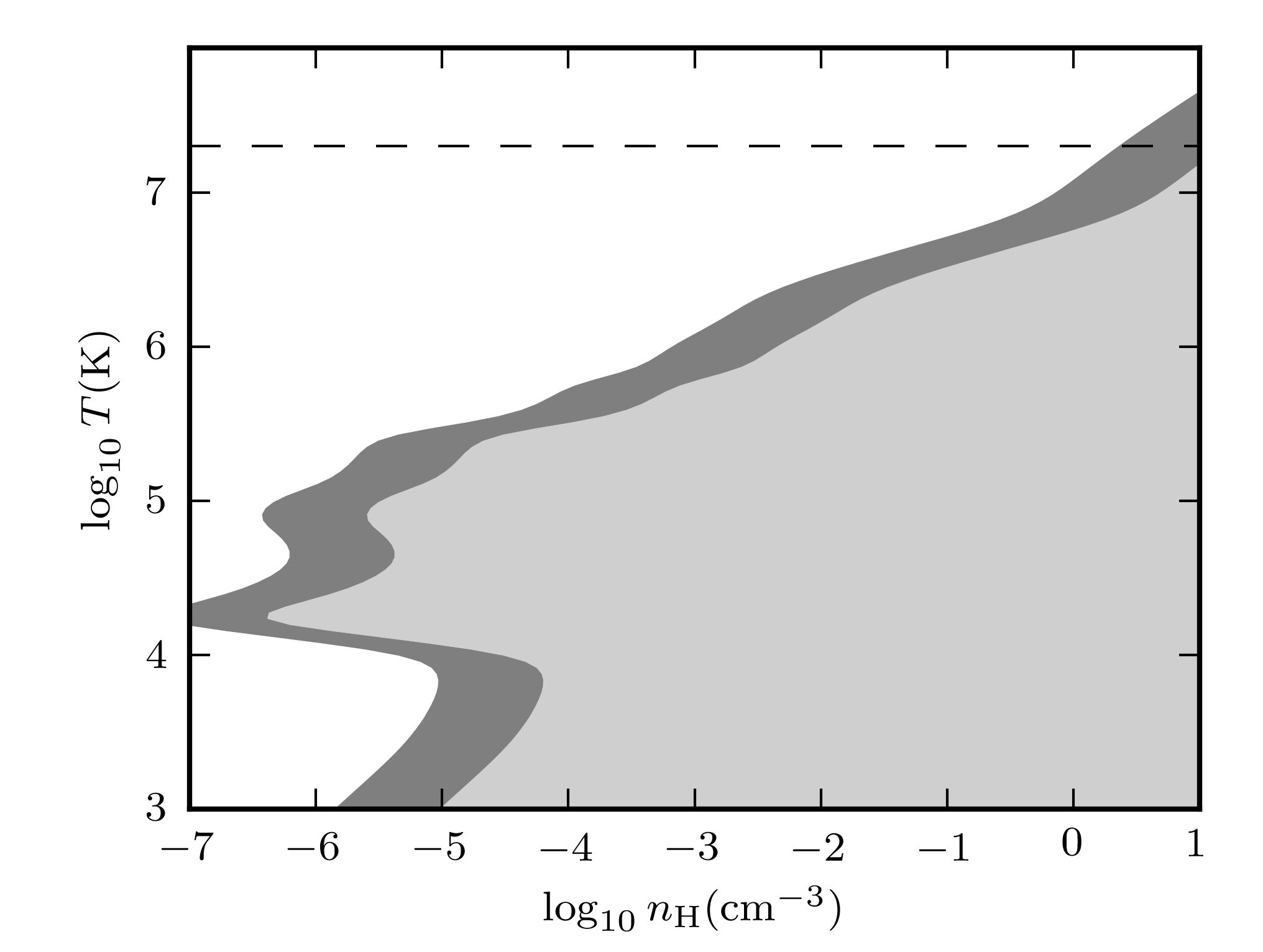}
\caption{Minimum re-heating temperature $T$ required to avoid numerical overcooling as a function of hydrogen number density $n_{\rm H}$, assuming an SPH resolution of $m_{\rm SPH}=10^6 M_\odot$ and solar metallicities. Cooling is so rapid in the {\em shaded} regions that the transition from thermally driven to momentum driven expansion phases of supernova bubbles is so fast that much of the injected energy will be lost to radiation. The \emph{light grey shaded region} corresponds to Eq.~(\ref{eq:sneCrit2}), the \emph{dark grey shaded} is the more demanding Eq.~(\ref{eq:sneCrit}). The \emph{white region} is where the reheating temperature is sufficient to force thermal feedback despite resolution concerns. \emph{Dashed line} is an estimate of the specific energy of SNe from \citet{Kay_et_al_02}.}
\label{fig:supernovae}
\end{figure}

In Fig.~\ref{fig:supernovae} we explore the parameter space for modelling thermal feedback in an SPH simulation with $10^6 M_\odot$ particles. At each density, there is a minimum temperature required to drive an adiabatic blast wave phase. The light grey region is defined by the sound crossing time argument of Eq.~(\ref{eq:sneCrit2}) and the dark grey is from Eq.(\ref{eq:sneCrit}). In the white region we expect effective thermal feedback; in the dark grey region it will be suppressed by overcooling in shocks, and finally in the light grey the code will be unable to produce a shock at all.

It is helpful to introduce some numbers. For gas with hydrogen number density $n_{\rm H} = 1$ cm$^{-3}$, a $m_{\rm SPH}=10^6 M_\odot$ particle would need to be heated to a temperature of $T_s\approx 5 \times 10^6$K to be in the pressure driven phase, according to Eq. (\ref{eq:sneCrit2}). However our resolution study suggests that we need a higher temperature of $\approx 10^7$K to prevent excessive overcooling through the shock, implying that at this resolution the simulation cannot properly represent the effects of thermal feedback. Note also from Eq.~(\ref{eq:sneCrit}), that this improves only very slowly with improved resolution, $\propto m_{\rm SPH}^{2/9}$. Relating back to models of feedback, this is somewhat problematic as the specific energy of supernovae \footnote{The specific energy of supernovae is the energy released by supernovae per unit mass of star-forming gas \citep{Kay_et_al_02}, i.e. if all supernovae were to ignite simultaneously we would reach this mean temperature.} is estimated to only be around $2 \times 10^7$K (dashed line in Fig.~\ref{fig:supernovae}) which can still be too low for the simulation code to properly follow the thermal evolution of the explosion.

Clearly this has important consequences for prescriptions for supernova-driven thermal feedback. In densities above $n_{\rm H} = 1 \, \rm cm^{-3}$ the thermal feedback starts to reduce its effectiveness, even if we ignite all our SNe in a single timestep, yet this is one of the key environments where feedback is required. 

One way to drive feedback at this resolution, whilst still maintaining a globally consistent initial mass function, is to stochastically inject the energy of SNe due to star formation. Since the mean thermal energy of a single particle can then be greater than $2 \times 10^7$K we can remain in the effective region of Fig. \ref{fig:supernovae}. To balance the IMF other star forming particles will need to receive less or no SNe energy. The alternative is to increase the resolution, but again the low exponent of $m_{\rm SPH}$ in Eq. (\ref{eq:sneCrit}) makes this quite prohibitive. 

The issue is further complicated by the existence of a multi-phase ISM not resolved by the simulation. This is the motivation for many of the prescriptions for feedback, such as applying a fraction of the supernovae energy as kinetic energy \citep{Navarro_White_93}, disabling the cooling of thermal bubbles \citep{Gerritsen_97} or releasing the energy of many accumulated supernovae in one step. A more thorough discussion of all these methods can be found in \citet{Kay_et_al_02}. Another approach is to model the net feedback effects by a subgrid model (for example an imposed equation of state without cooling as in \citealp{Schaye2010}), or to model the hot and cold phases separately by representing clouds in the cold phase as collisionless particles \citep{Booth2007}.

At lower densities it becomes easier to thermally drive a blast wave due to the reduced cooling rate, reinforcing the importance of simulating a multiphase ISM. For their star formation threshold of $n_{\rm H} = 10^{-1}$ cm$^{-3}$, the high-resolution {\sc OWLS} simulations of \cite{Schaye2010} can represent thermally driven SNe at the edges of discs, but not in more central regions. Indeed, at higher densities the required temperatures rapidly reach extreme values. \cite{Booth_and_Schaye_09} note that in their simulations, AGN feedback requires reheating temperatures $T_s > 10^8 $K, as at lower temperature the energy is simply radiated away. We believe that this problem is not a physical one but one of resolution. At lower temperatures the density is higher, the cooling faster, and the cooling region behind the shock cannot be resolved, as is clear from Fig.~\ref{fig:supernovae}.

\subsection{Shocks at the sound speed}
As an interesting aside it is worth considering that there will usually be an upper limit to the resolution required. If we assume that the weakest shock has a velocity on the order of the sound speed, $c_s = (\gamma (\gamma -1)u)^{1/2}, $ then the minimum requirement for the particle mass for a given problem will be
\begin{equation} 
m_{\rm SPH} = (0.08)^3 \left(\gamma (\gamma-1) \right)^{9/2} \,\min_{x \in \mathrm{V}} \left\{  |\dot{u}_\Lambda|^{-3} u^{9/2}  \rho \right\}\,.
\end{equation}

Unfortunately such a limit will usually be very small indeed, at least for cosmological simulations, because of the low sound speed of cold, dense gas present in galactic discs. However if one chooses to go down this path, then one can examine the following criteria. If we have a conventional collisional cooling function then $\dot{u}_\Lambda / \rho$ is independent of density, and we can make the additional assumptions that $\dot{u}_\Lambda \to 0$ as $u \to 0$, and for large $u$
\begin{equation} 
\frac{\left|\dot{u}_\Lambda \right| }{ \rho} \sim u^{1/2}
\end{equation}
(thermal bremsstrahlung), giving 
\begin{equation} 
m_{\rm SPH} \propto  \min_{x \in \mathrm{V}} \left\{  \rho^{-2} \right\}\,,
\end{equation}
{\em i.e.} the smallest particle mass is determined by the highest density in the problem. This analysis is of course not valid with Compton cooling via the CMB, or the presence of a UV background, as neither process is collisional.

\section{Conclusions}
In this paper we have examined the role of radiative cooling in shocks. We have found a general analytical solution for 1d piecewise linear collisional cooling functions and compared it to numerical simulations of the same shock, performed with an SPH code ({\sc Gadget}) and an AMR code ({\sc FLASH}). These codes smear out the shock over several particles or cells, and such an artificial \lq pre-shock\rq\ results in numerical overcooling which may prevent the formation of a hot post-shock region. We have estimated a general resolution criterion to avoid such overcooling, and applied it to the problems of virial shocks and the production of hot gaseous haloes. We have found that to avoid numerical overcooling of accretion shocks onto haloes that should develop a hot corona requires a particle or cell mass resolution of $10^6 M_\odot$ (Fig. \ref{fig:virial_uvback}), which is within reach of current state-of-the-art simulations.

Similarly, we have applied our estimates to thermal feedback from AGN or supernovae blast waves, in the presence of radiative cooling. We have seen that the energy required to drive thermal feedback at a given mass scale, for current numerical results, is an order of magnitude higher than one would expect just from physical considerations.  For cosmological simulations ($10^6 M_\odot$ gas particles) of an $n_{\rm H}=1.0 \, {\rm cm}^{-3}$ interstellar medium we see (Fig. \ref{fig:supernovae}) that temperatures in excess of $10^7$~K are required to effectively drive thermal feedback by avoiding spurious suppression of the feedback by numerical overcooling.

Although all of these issues can be rectified by increasing the resolution, the minimum thermal energy of injected feedback required to avoid artificial cooling scales weakly 
with decreasing particle mass, $\propto m_{\rm SPH}^{2/9}$, see Eq.~(\ref{eq:sneCrit}). Consequentially, a potentially fertile region of study may be that of cooling switches, {\em i.e.} a criterion for disabling cooling through a shock. Such a switch would allow a simulation to resolve temperatures much closer to the physical temperatures of radiative shocks without requiring extreme resolutions. Unfortunately it is not a straightforward problem to have a criterion that will consistently suppress cooling in the presence of shocks yet does not affect cooling in regions where there are no shocks. Since we can never hope to completely remove resolution effects it seems sensible to have a more limited aim, perhaps to capture the temperatures of shocks up to some maximum cooling rate. As such one might wish to suppress cooling, when the cooling time is greater than some fraction of the shock heating time.  We intend to explore this avenue in a further paper.

Further work could include the effects of shock-induced non-collisional ionizational equilibrium (CIE) or non-thermalised gas. Since the resolution can make such a significant modification to the thermal history of a gas, we expect a criterion due to non-CIE may be quite strict.

\section*{Acknowledgements}   
Peter Creasey would like to acknowledge the support of an STFC studentship. The authors would like to thank Rob Crain for the re-use of simulation data, and Volker Springel for the provision of the SPH code \gadget. The authors thank Peter Thomas, Frazer Pearce, Justin Read, Tom Abel, Daniel Price, Romain Teyssier and the anonymous referee for encouragement and very useful suggestions. Some of the calculations for this paper were performed on the ICC Cosmology Machine, which is part of the DiRAC Facility jointly funded by STFC, the Large Facilities Capital Fund of BIS, and Durham University. The \flash\ software used in this work was in part developed by the DOE-supported ASC/Alliance Center for Astrophysical Thermonuclear Flashes at the University of Chicago.

\bibliographystyle{mn2e}
\bibliography{bibliography} 
\bsp

\begin{appendix}
\medskip
\section{Radiative shocks with piecewise linear cooling functions}
\subsection{Similarity solution for a 1d radiatively cooling shock} 
\label{sec:Piecewise}

We start with an ideal gas with adiabatic index $\gamma$
\begin{equation}
p = (\gamma - 1) \rho u\,,
\end{equation}
and a collisional radiative cooling function
\begin{equation}
\left. \mathrm{d}u \right|_\Lambda = -\rho f(u)  \mathrm{d}t\,.
\end{equation}

These combine to give an evolution of
\begin{equation}
\mathrm{d}u = (\gamma - 1) \frac{u}{\rho} \mathrm{d}\rho - \rho f(u) \mathrm{d}t\,.
\end{equation}

Stationary solutions of a post shock cooling region satisfy integrals of the mass and momentum equations, i.e.
\begin{eqnarray}
\rho (v - u_s) &=& \rho_0 (v_0 - u_s) \\
p + \rho (v-u_s)^2 &=& p_0 + \rho_0 (v_0 - u_s)^2\,,
\end{eqnarray}
where  $u_s$ is the shock velocity and $\rho_0,p_0, v_0$ denote the density, pressure and velocity at some arbitrary downstream point. Thus the density, velocity and thermal energy can be written in terms of a similarity variable $\lambda$

\begin{eqnarray}
\rho / \rho_0 &=& \lambda \\
\frac{v-u_s}{v_0 - u_s} &=& \lambda^{-1} \\
u / u_0 &=&  (a+1) \lambda^{-1} - a \lambda^{-2}\,,
\end{eqnarray}
with
\begin{equation}
a = \frac{\rho_0 (v_0 - u_s)^2}{p_0}\,.
\end{equation}

Now we assume we have a piecewise linear cooling function, i.e. we can solve each segment separately with the linear cooling function
\begin{equation}
\label{eq:lincool}
f(u) = A (u-u_c)\,,
\end{equation}
where $A$ is some constant and $u_c$ denotes the `cold' thermal energy where cooling vanishes. This gives an o.d.e for $x$ of the form
\begin{equation}
\frac{\mathrm{d}x}{\mathrm{d}\lambda} = \frac{v_0-u_s }{A \rho_0} \left( \frac{\gamma (a+1) \lambda^{-4} - (\gamma + 1) a \lambda^{-5}}{(a+1) \lambda^{-1} - a \lambda^{-2} - u_c/u_0} \right)\,,
\end{equation}
which can be solved generally, however in the case of $u_c = u_0$ we have the particularly simple case,
\begin{eqnarray*}
x - x_0 &=& \frac{v_c-u_s }{A \rho_c}\left[\frac{\gamma-a}{a-1} \log (\lambda^{-1}-1) + \right. \\
& &  \frac{1-a\gamma}{(a-1)a^2} \log (1 - a\lambda^{-1}) \\
& & \left. - \frac{a+1}{a} \lambda^{-1} - \frac{\gamma+1}{2}
  \lambda^{-2}  \right] \\
\lambda &\in & \left[\frac{a}{a+1}\frac{\gamma+1}{\gamma},1\right]\,,
\end{eqnarray*}
the left hand limit for $\lambda$ coming from entropy considerations. An example cooling shock of this form can be seen in Fig. \ref{fig:analytic}.

\subsection{Colliding gas}
\label{sec:RHShock}
Assume two homogeneous flows collide from the left and right, with properties $\rho_0, p_0, \pm v_0$. With no cooling a hot, static region is created in the centre, with properties $p_c$, and $\rho_c$. The mass, momentum and energy equations are
\begin{eqnarray}
(v_0 - u_s) \rho_0 & = & -\rho_c u_s \\
(v_0 - u_s)^2 \rho_0 + p_0 & = & \rho_c u^2_s + p_c \\
\frac{p_0}{\rho_0} + \frac{1}{2} (\gamma - 1) v_0^2 & = & \frac{p_c}{\rho_c}\,,
\end{eqnarray}
where $u_s$ is the velocity of the left moving shock in the rest frame. Eliminating $p_c$, $\rho_c$ gives
\begin{equation}
u_s^2 + \frac{1}{2} (\gamma - 3) u_s v_0 = \frac{p_0}{\rho_0} + \frac{1}{2} (\gamma - 1) v_0^2\,,
\end{equation}
so
\begin{equation}
u_s =  -\frac{1}{4} (\gamma - 3) v_0 - \frac{1}{4} \sqrt{ v_0^2  (\gamma + 1)^2 + 16 \frac{p_0}{\rho_0}}\,.
\end{equation}

Assume now that there {\it is} cooling and that the gas in the centre cools to the temperature of the pre-shock gas (where cooling is assumed to vanish). In this case the mass, momentum and energy equations are
\begin{eqnarray}
(v_0 - u_s) \rho_0 & = & -\rho_c u_s \\
(v_0 - u_s)^2 \rho_0 + p_0 & = & \rho_c u^2_s + p_c \\
p_0 / \rho_0 & = & p_c / \rho_c\,,
\end{eqnarray}
where these equations are only dependent on the cooling function via the thermal state at which cooling vanishes, $p_0 / \rho_0$. Eliminating $p_c$, $\rho_c$ gives
\begin{equation}
u_s^2 \rho_0 - u_s v_0 \rho_0 - p_0 = 0\,.
\end{equation}
The solution for the shock velocity $ u_s = v_0 / 2 -
\sqrt{(v_0/2)^2 + p_0 / \rho_0}$. $p_c, \rho_c$ can be found by
substitution.

The conditions immediately to the right of the shock ($v_s$, $\rho_s$,
$p_s$) can be found from the usual Rankine-Hugoniot relations,
\begin{eqnarray}
(v_0 - u_s) \rho_0 & = & (v_s - u_s) \rho_s \\
(v_0 - u_s)^2 \rho_0 + p_0 & = & (v_s - u_s)^2 \rho_s + p_s \\
\frac{1}{2} (v_0 - u_s)^2  + \frac{\gamma}{\gamma-1} \frac{p_0}{\rho_0} & = & \frac{1}{2} (v_s - u_s)^2  + \frac{\gamma}{\gamma-1} \frac{p_s}{\rho_s}\,.
\end{eqnarray}

\end{appendix}

\label{lastpage}
\end{document}